\documentclass[twocolumn]{aastex7}

\usepackage{amsmath}
\usepackage{gensymb}
\usepackage{booktabs}

\newcommand{\hu}[0]{\hat{h}}
\newcommand{\hsat}[0]{h_{\rm sat}}
\newcommand{\CAPE}[0]{{\rm CAPE}}
\newcommand{\paperedit}[1]{#1}

\shorttitle{Wind shear and convection}
\shortauthors{Sankar et al.}

\graphicspath{{./}{figs/}}

\begin{document}

\title{Wind shear and the role of eddy vapor transport in driving water convection on Jupiter}

\author[0000-0002-6794-7587]{Ramanakumar Sankar}
\affiliation{Center for Integrative Planetary Sciences \\
University of California, Berkeley,
Berkeley, CA, USA}
\email[show]{ramanakumar.sankar@berkeley.edu}

\author[0000-0003-2804-5086]{Michael H. Wong}
\affiliation{Center for Integrative Planetary Sciences \\
University of California, Berkeley,
Berkeley, CA, USA}
\email{mikewong@ssl.berkeley.edu}

\author[0000-0001-6111-224X]{Csaba Palotai}
\affiliation{Department of Aerospace, Physics and Space Sciences \\
Florida Institute of Technology,
Melbourne, FL, USA}
\email{cpalotai@fit.edu}

\author[0000-0002-3669-0539]{Shawn Brueshaber}
\affiliation{Michigan Technological University, \\
Mechanical Engineering-Engineering Mechanics,\\
1400 Houghton, MI 49331 USA}
\email{srbruesh@mtu.edu}









\begin{abstract}
Recent observations of convection in the jovian atmosphere have demonstrated that convection is strongly concentrated at specific locations on planet. For instance, observations of lightning show that the cyclonic features (e.g,. belts and folded filamentary regions - FFRs) show increased convective activity compared to anti-cyclonic regions. Meanwhile, the distribution of ammonia and water vapor show a large enrichment near the equator, which is also suggestive of strong upwelling and convective activity. Marrying these different observations is challenging due to a lack of data concerning the characteristics of the deep jovian atmosphere, and a resulting inability to observe the true deep source of the various convective phenomena. To understand the nature of these convective events and \paperedit{the role of the } structure of the deep atmosphere \paperedit{in driving convective events}, we run simulations of cloud formation and convection using the Explicit Planetary hybrid-Isentropic Coordinate General Circulation Model (EPIC GCM). We vary the dynamics of the atmosphere by parameterizing the deep wind shear and studying the resulting effect on the strength, frequency and distribution of convective storms. We find that convection in our model is strongly tied to the local dynamics and the deep wind shear. We further decompose the generation of convective available potential energy (CAPE) into three components (thermal, mechanical, and moist/chemical), and find that the chemical mechanism is the strongest component, working to advect water vapor from moisture-rich regions to moisture-poor regions and to drive convection along a ``moisture front.'' 
\end{abstract}



\section{Introduction} \label{sec:intro}
Jupiter's energy balance poses many questions about the processes that drive the dynamics of the atmosphere and cloud formation on the planet. Specifically, while moist convection is thought to be both a vital mechanism in transferring heat vertically \citep{Gierasch2000} and in driving atmospheric flows \citep{Ingersoll2000, Showman2007}, the process itself is poorly understood, specifically due to the lack of comprehensive observations \citep[e.g.,][]{Hueso2022} \paperedit{and the difficulty in simultaneously modeling the coupled effects of convection, cloud microphysics, radiative transfer and fluid dynamics \citep{Palotai2022} even though individual effects of moist convection have been separately modeled with various parameterizations \citep{Showman2007,LianShowman2010,Sankar2022,Ge2024,Boissinot2024}}. Particularly, the diversity of convective features is not well constrained and is instead inferred from multi-modal observations such as lightning \citep{Brown2018}, distribution of volatiles \citep{Li2018,Guillot2020b,Moeckel2023} and zonal disturbances \citep{Fletcher2017,Sankar2022,Hueso2022}.

One of the key issues is in understanding this diversity and interpreting why specific regions are conducive of convection (or more accurately, show signatures of convective activity), while other don't. For example, belts generally show more lightning compared to zones \citep{Brown2018}, and Folded Filamentary Regions (FFRs) and other cyclonic structures show convective features and lightning more than anti-cyclones \citep{Vasavada2005}. To a first-order explanation, cyclonic features on Jupiter generally are more unstable due to an expansion of the isentrope near their roots \citep{DowlingGierasch1989}, thereby explaining why belts, which have cyclonic shear, generally have more convective activity. However, distributions of volatiles and aerosol morphologies show that we should expect upwelling in the zones to produce high altitude clouds, and downwelling in the belts \citep[see][and references therein]{dePater2023}, which is at odds with the dynamical instability paradigm. 

Recent observations using microwave and radio instruments show weak correlation between the brightness temperature at depth and the cloud top zonal wind profiles \citep{dePater2019,Fletcher2021}. While this could be indicative of deep zonal shear, the values obtained from inverting the thermal wind equation at depth results in uncharacteristically large values \citep{Fletcher2021}. An equally likely interpretation could suggest variability in the concentration of ammonia, sustained by circulation within multiple stacked Ferrel-like cells \citep{Fletcher2020JGR,Duer2021}. A comprehensive understanding of convection on Jupiter that successfully meshes these observations with the aforementioned instability criteria is still missing. 

In this study, we use the Explicit Planetary hybrid-Isentropic Coordinate General Circulation Model \citep[EPIC GCM, ][]{Dowling2006} to simulate convective cloud formation and investigate the dynamics of convective storms on Jupiter. We use the cloud microphysics parameterization \citep{Palotai2008} and the Relaxed Arakawa-Schubert \citep{Moorthi1999,Sankar2022} convective scheme to simulate sub-grid scale moist convection in an effort to bridge the various observations of convection on Jupiter. Particularly, for this study, we focus on the effect of the vertical wind shear, which strong affects the stability of features, and eddy momentum and mass transport in the atmosphere \citep[see e.g.,][]{GarciaMelendo2005}. We detail an overview of our method to diagnose the effects of dynamics on convection in Section~\ref{sec:CAPE}, discuss our methods in Section~\ref{sec:methods}. We present our results in Section~\ref{sec:results} and discuss these findings within the limitations of our model and in the context of known observations of convection in Section~\ref{sec:discussion}.

\section{A review of convection and convective triggers}
\label{sec:CAPE}
Convection in the atmosphere is through the generation of buoyant instability. Essentially, convection from the cloud base ($z_{\rm base}$) to the cloud top ($z_{\rm top}$) occurs when the CAPE, given by,
\begin{equation}
    \CAPE = \int_{z_{\rm base}}^{z_{\rm top}} g \dfrac{\hat{T_v} - T_v}{\hat{T_v}} \; dz,
\end{equation}
is greater than a critical value $\CAPE_{\rm crit}$, which governs the stability of the atmosphere to small perturbations \citep{ArakawaSchubert1974}. $T_v$ is the virtual temperature, i.e., the temperature of a parcel of dry air with the same density as the moist air, and quantities with the hat represent those within the updraft (e.g., $\hat{T_v}$ is the virtual temperature of the convecting parcel) while those without the hat represent the properties of the ambient atmosphere. This can be rewritten as,
\begin{equation}
    \label{eq:buoyancy}
    \CAPE \sim \int_{z_{\rm base}}^{z_{\rm top}} \dfrac{g}{\tilde{L}} \left(\hat{h} - \hsat\right) \; dz,
\end{equation}
 where $g$ is gravitational acceleration, $\tilde{L}$ is \paperedit{a term that scales the latent heat with virtual effects} \citep{Moorthi1999}. $\hat{h}$ is the moist static \paperedit{energy} of an updrafting parcel, given by,
\begin{equation}
    \hat{h} = \hat{H} + \hat{\Phi} + L_v \hat{q},
\end{equation}
where $\hat{H}$ is the enthalpy, $\hat{\Phi}$ is the geopotential, $L_v$ is the latent heat and $\hat{q}$ is the \paperedit{vapor} mass mixing ratio of the condensible species \citep{Moorthi1999,Sankar2022}, all within the updrafting parcel. $\hat{h}$ is conserved during moist ascent, while entraining dry air generally reduces $\hat{h}$. When this is higher than the atmospheric saturated static \paperedit{energy}, $\hsat$, where,
\begin{equation}
    \hsat = H + \Phi + L_v q_{\rm sat},
\end{equation}
and $q_{\rm sat}$ is the saturated \paperedit{vapor} mass mixing ratio of the condensing species, the stability of the atmosphere decreases, resulting in convection. Note that $\hsat$ is defined as the atmospheric quantity and therefore, describes the ambient atmosphere outside the convective parcel. In large scale convection, when the large scale forcing \paperedit{(i.e., with respect to the numerical model, this defines grid scale or larger, while convective is generally sub-grid scale)} from the dynamics increases the bouyancy (and thereby reducing stability), the response of the atmosphere is through convection, which creates a moist convective energy flux that balances the rate of change of buoyancy. Therefore, the moist convective flux is given by,
\begin{equation}
    F_{\rm MC} \sim \dfrac{d(\CAPE)}{dt}.
\end{equation}

As such, in order to understand the source and strength of convection in the atmosphere, we need to understand the processes that drive an increase in $\CAPE$ as given by Equation~\ref{eq:buoyancy}. 
We can write,
\begin{equation}
    \dfrac{d\hu}{dt} = C_p \dfrac{d\hat{T}}{dt} + \dfrac{d\hat{\Phi}}{dt} + L_v \dfrac{d\hat{q}}{dt},
\end{equation}
and similarly, 
\begin{equation}
    \dfrac{d\hsat}{dt} = C_p \dfrac{dT}{dt} + \dfrac{d\Phi}{dt} + L_v \dfrac{dq_{\rm sat}}{dt}.
\end{equation}
Combining this, we get,
\begin{align}
    \begin{split}
        \dfrac{d(\CAPE)}{dt} \sim \int_{z_{\rm base}}^{z_{\rm top}}  & \dfrac{g}{\tilde{L}} \left[ C_p \left(\dfrac{d\hat{T}}{dt} - \dfrac{dT}{dt}\right) \right. \\*
        & + \left( \dfrac{d\hat{\Phi}}{dt} -  \dfrac{d\Phi}{dt} \right) \\* 
        & \left. + L_v \left(\dfrac{d\hat{q}}{dt} - \dfrac{dq_{\rm sat}}{dt}\right) \right] \; dz
    \end{split}
\end{align}

To simplify the three terms, which now depend on an updraft parcel value and an atmospheric value, we can assume for simplicity that the updraft does negligible exchange of mass with the atmosphere (i.e, very little entrainment or detrainment). Note that this is not how the convection is treated in our model, which features a more complete formulation of the updraft entrainment profile, but is instead a reasonable assumption for strong updrafts to simplify the conceptual interpretation in this section. Therefore, we can assume that $\hu = h(z_{\rm base})$, which gives,

\begin{align}
    \begin{split}
    \dfrac{d(\CAPE)}{dt} \sim \int_{z_{\rm base}}^{z_{\rm top}} \dfrac{g}{\tilde{L}} & \left[ 
    \underbrace{C_p \left(\dfrac{dT_{\rm base}}{dt} - \dfrac{dT}{dt}\right)}_{\text{Thermal}} \right. \\*
    & + 
    \underbrace{\left( \dfrac{d\Phi_{\rm base}}{dt} -  \dfrac{d\Phi}{dt} \right)}_{\text{Mechanical}} \\* 
    & \left. + 
    \underbrace{L_v \left(\dfrac{dq_{\rm base}}{dt} - \dfrac{dq_{\rm sat}}{dt}\right)}_{\text{Chemical}} \right]
    \end{split}
\end{align}\label{eqn:decomposed}

These three terms represent three different physical processes in the atmosphere, which we dub the thermal, mechanical and chemical tendencies, respectively. We detail these terms below. 

\subsection{Thermal effects: direct heating/cooling}
The first term is given by the rate of the change of temperature. This is primarily through advection or from direct cooling/heating (e.g., from latent heat release or insolation). Therefore, this term corresponds to thermal processes \paperedit{that modify the internal energy of the atmosphere and } drive convection.

\subsection{Mechanical effects: atmospheric thickness changes}
\paperedit{The second term corresponds to changes in the geopotential in the atmosphere which }corresponds to the mechanical forcing (i.e., squeezing or expanding a column of atmosphere, which drives or inhibits convection). \paperedit{Increasing the geopotential height between the cloud base and a layer above increases the potential energy between a parcel of air in that layer and the cloud base.}

\subsection{Chemical effects: changes in moisture}
Finally, the last term corresponds to changes in the moisture mixing ratio within the column. This can be through microphysical effects (such as sublimation of cloud ice/snow) or through advection of vapor (e.g., convergence of vapor to create moisture rich regions or divergence to create dry spots). This term corresponds to the chemical mechanisms that drive convection \paperedit{by virtue of changing the effect of latent heat release in driving and maintain convection}.

Therefore, the combination of these three processes (thermal, mechanical and chemical) drive convection in the atmosphere. In this study, our goal is to understand how these processes are driven by the global atmospheric flows on Jupiter. Specifically, we determine whether specific locations where convection is observed is dominated by one process, as opposed to the other two.

\section{Methods} \label{sec:methods}

In this study, we use the EPIC model to simulate the jovian atmosphere and study cloud formation and moist convection. 

\subsection{Model setup}
\begin{figure}
    \centering
    \includegraphics[width=\columnwidth]{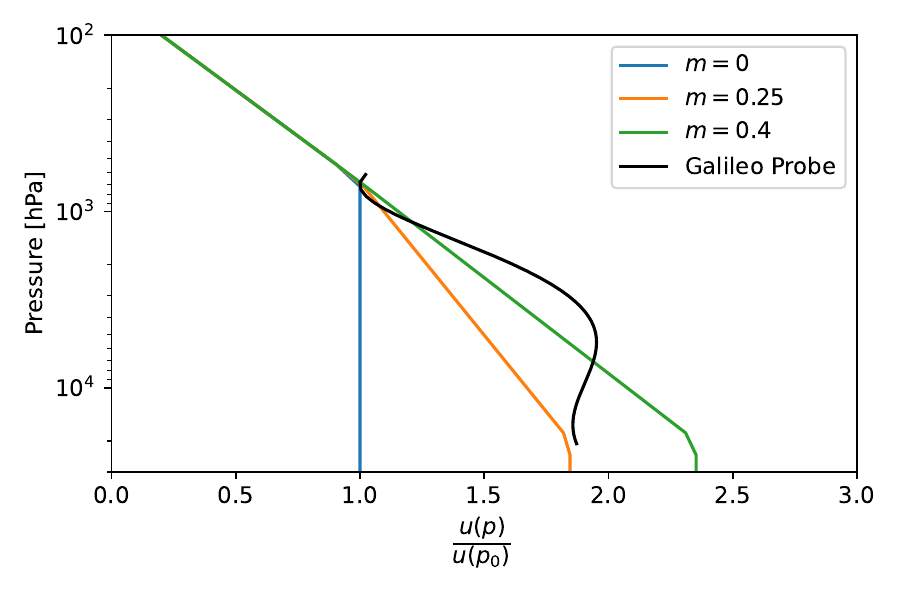}
    \caption{Shear profile for different values of $m$ used in the model.}
    \label{fig:shear_profile}
\end{figure}

To study the effect of wind shear on the atmosphere, we parameterize the vertical wind profile similarly to \citet{GarciaMelendo2005}, who define a wind shear slope $m$, in log-pressure space. Our wind profile is of the form,
\begin{equation}
    u(p) = 
    \begin{cases} 
        u(p_0) \left(1 - \dfrac{1}{2.4} \log\left(\dfrac{p}{p_0}\right)\right) & p < p_0, \\
        u(p_0) \left(1 + m \log\left(\dfrac{p}{p_0}\right)\right) & p_0 < p < p_c, \\
        u(p_0) \left(1 + m \log\left(\dfrac{p_c}{p_0}\right)\right) & p > p_c,
    \end{cases}
\end{equation}
where $p_0$ is the location of the cloud top \paperedit{(diagnosed from Hubble CH$_4$ I/F, as detailed below)} and is variable with latitude, $u(p_0)$ is the cloud-tracked wind speed at $p_0$ and $p_c = 30$ bar, is the ``critical" pressure in our model where the wind shear goes to 0. An illustration of this profile is shown in Figure~\ref{fig:shear_profile} for the different values of $m$ tested in this study. For comparison, the Galileo Probe Doppler Wind Experiment revealed a wind shear roughly corresponding to $m = 0.95$, but limited to a shallower shear layer with $p_c$ = 3.5 bar \citep{atkinson++1998}. It is plausible that $m$ varies horizontally and vertically in the real Jovian atmosphere, but in the absence of relevant measurements, we simply ran simulations using constant $m$ globally to test the effect of shear on the convective process.

\paperedit{Our parameterizations consider only shallow regions of the atmosphere (depth of $\sim$130 km at 20 bar), compared to wind retrievals at depths of more than 1000 km from inversion of gravity field data. For example, \citet{Cao2023} find winds at 1500 km depth with magnitudes more than twice as large as the cloud-tracked winds. Our test cases are thus within the range of observational constraints from the Galileo Probe and Juno gravity data, especially if vertical shear is minimal between the 100 km and 1500 km levels.}

\begin{figure}
    \centering
    \includegraphics[width=\columnwidth]{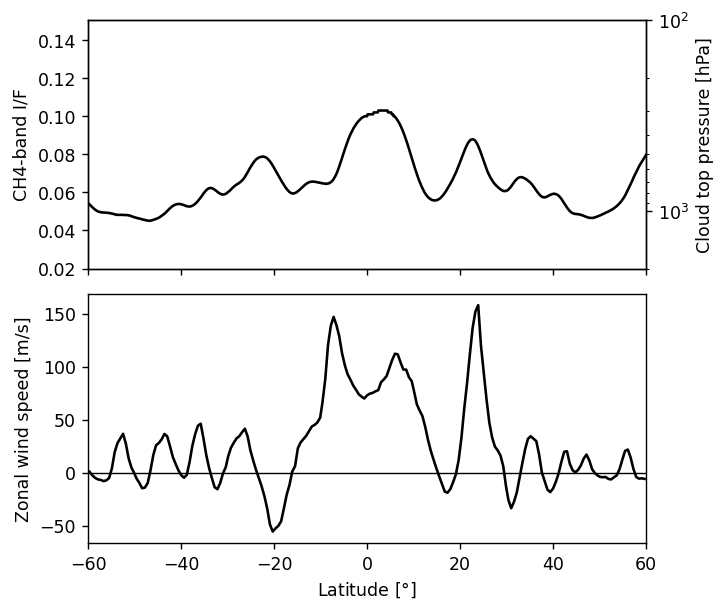}
    \caption{Top: I/F from CH4-band (889 nm) Hubble observations and the corresponding cloud top location used in our wind initialization. Bottom: The input zonal wind speed applied at the corresponding pressure level. Both from \citet{Tollefson2017}.}
    \label{fig:zonal_wind}
\end{figure}
\begin{figure}
    \centering
    \includegraphics[width=\columnwidth]{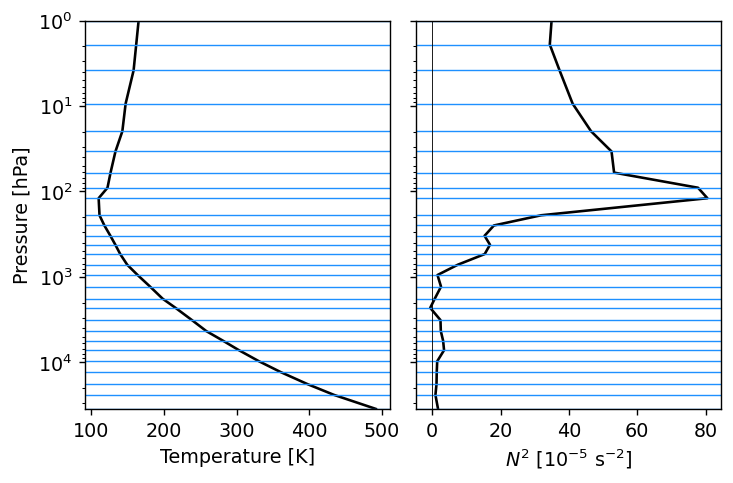}
    \caption{Input temperature (left) and static statibility parameter (right). The model's vertical layers are shown with the blue lines. We apply this profile at a latitude of $23\degree$ N, which we found to produce the most stable configuration.}
    \label{fig:tp}
\end{figure}

For all our simulations, our model extended from $-60\degree$ to $60\degree$ planetographic latitude, with 200 grid points, and from $0$ to $120\degree$ longitude with $256$ grid points, which is a significantly expanded domain compared to previous simulations of convection using EPIC \citep{Sankar2022}. This results in a roughly $0.5\degree \times0.5 \degree$ resolution, which is just under $800$ km horizontal resolution meridionally and between 300-600 km zonally. 
This is comparable to recent studies of global jovian dynamics \citep{Young2019a}, but generally on the order of domain size for cloud-resolving models \citep{Hueso2001,Ge2024}. 
Vertically, we use 28 layers spaced unevenly between $0.1$ mb and $40$ bars (blue lines in Fig.~\ref{fig:tp}). 
Levels where clouds are expected to form have the highest resolution in our model ($\sim 5$ km), and is slightly larger than recent cloud-resolving models \citep[$\sim 2$ km,][]{Ge2024}. We apply an 8th-order hyperviscosity and divergence damping to remove high frequency modes. Their corresponding coefficients are $\nu_8 = (0.5) [(1/2400) \Delta y^8/\Delta t]$ m$^8$/s and $\nu_{\rm div} = (0.5) [(1/3) \Delta y^2/\Delta t]$ m$^2$/s. For our $m=0.4$ case we found that this does not stop high frequency modes, and use a 6-th order hyperviscosity with a value of $\nu_6 = (0.1) [(1/600) \Delta y^6/\Delta t]$ m$^6$/s. We find that there is no appreciable change in the dynamics of the atmosphere between $\nu_6$ and $\nu_8$ except for preventing spurious oscillations near the meridional boundaries.

The input zonal wind profile is from \citet{Tollefson2017}, which we apply to the cloud top location. 
We define a variable cloud-top location as a function of latitude, using 
zonally-averaged Hubble methane-band reflectivity (sensitive to high-altitude clouds/haze) as a proxy for cloud height.
Cloud top altitudes are hard to precisely constrain, so we 
simply interpret low reflectivity as deep clouds around 1 bar, and high reflectivity as high clouds at 200 mb. 
The resulting cloud top locations and the input zonal wind speed at this location is shown in Figure~\ref{fig:zonal_wind}. We then apply the shear profile defined above to create a 3D wind field. The variable cloud-top approach is supported by some of our simulation results, as discussed below in Sec.~\ref{sec:results_cloudformation}.

The input temperature profile from \citep{Moses2005} is applied at $24\degree$ N and we then use the thermal wind balance to obtain the 3D temperature field from the wind field. We tested different points to integrate the thermal wind balance, and found that this location produced the most stable configuration due to the presence of the steep gradient due to the strong jet. Ultimately, the model is not sensitive to this choice in the long term \paperedit{\citep[for a desciption of how this choice plays a role in the initialization, the reader is referred to Section 3.2 in][]{Dowling1998}}, but is instead only useful to ensure that the model does not crash during the spinup phase. \paperedit{Specifically, in some scenarios, choosing a different points is sensitive to the meridional gradients when integrating the thermal wind equation and we find that using this location produces the most consistent result.} The input temperature profile, and corresponding Brunt-V\"ais\"al\"a frequency, which we calculate using both the temperature gradient and mass loading effects \citep{Dowling2006}, are shown in Figure~\ref{fig:tp}.

The deep composition for the model atmosphere includes H$_2$, He, H$_2$O, and NH$_3$, with the assumed volume mixing ratios (VMRs) given in Table~\ref{tab:comp}. The table lists enrichment factors relative to solar or protosolar composition for a number of frequently used tables of reference composition. Other species not included (e.g., CH$_4$, H$_2$, Ne) are expected to have minimal contributions to the mean atmospheric molecular weight and specific heat.
We initialize our moisture variables using the cold-trap assumption for each vertical column by specifying the deep vapor mixing ratio below the saturation point and follow the $95\%$ relative humidity curve above.
In this study, we fix the initial volatile distribution to study solely the effect of the difference in the \paperedit{large-scale} dynamics on convection. While convection has been shown to strongly scale with the water abundance \citep{Hueso2001,Sugiyama2014,Sankar2022} or be limited by the steep molecular weight gradient for large abundances \citep{Guillot1995,Li2015,Leconte2017}, we do not test the effect of water abundance here, and instead leave it to a future study.

\begin{table*}
\centering
\caption{Base composition for model atmosphere.}\label{tab:comp}
\begin{tabular}{wl{2cm} wc{3cm} wc{1.5cm} wc{1.5cm} wc{1.5cm} wc{1.5cm} wc{1.5cm} }
\hline
  &  &     \multicolumn{5}{c}{Relative to:}\\ \cline{3-7}
Species & VMR &  AG1989$^{a}$ & GS2001$^{b}$ & A2009$^{c}$ & L2010$^{d}$ & L2021$^{e}$  \\
\hline
H$_2$ & 0.859 & 1.0 & 1.0 & 1.0 & 1.0 & 1.0 \\
He & 0.157 & 1.0 & 1.1 & 1.0 & 1.0 & 1.0 \\
NH$_3$ & 5.63$\times 10^{-4}$ & 3.0 & 4.0 & 4.5 & 4.1 & 3.9 \\
H$_2$O & 4.33$\times 10^{-3}$ & 3.0 & 3.8 & 4.8 & 4.3 & 3.9 \\
\hline
\multicolumn{7}{l}{
  $^{a}$ Solar abundance table of \citet{Anders1989}, used in \citet{Wong2004}, }\\
\multicolumn{7}{l}{ \citet{Niemann1998}, and in EPIC4 native parameterizations. }\\
\multicolumn{7}{l}{  $^{b}$ Solar abundance table of \ \citet{Grevesse2001}, available in EPIC5 parameterizations. }\\
\multicolumn{7}{l}{ $^{c}$ Protosolar abundance table of \ \citet{Asplund2009}, used in \citet{Li2017,Li2024}, }\\
\multicolumn{7}{l}{ \citet{Ge2024}, and \citet{Moeckel2023}. }\\
\multicolumn{7}{l}{
  $^{d}$ Protosolar abundance table of  \citet{Lodders2010}, used in \citet{Leconte2017}.}\\
\multicolumn{7}{l}{ $^{e}$ Protosolar abundance table of  \citet{Lodders2021}, includes updated 3D non-LTE solar }\\
\multicolumn{7}{l}{ photospheric retrievals. }
\end{tabular}
\end{table*}

\subsection{Moist convective parameterization}
In this study, we use the Relaxed Arakawa-Schubert (RAS) module, which is a recent addition to the EPIC GCM \citep{Sankar2022}, to parameterize sub grid-scale water convection, to account for the coarse vertical and horizontal resolution in GCMs, which prevents full treatment of small scale convection. The RAS is a convective-adjustment scheme, which works by solving for a spectrum of convective clouds, each with a different vertical extent from a common cloud base, which satisfy the total convective energy budget within a column of the model atmosphere. The RAS module solves the vertical continuity equation for each cloud,
\begin{equation}
    \dfrac{\partial \eta}{\partial z} = E - D,
\end{equation}
where $\eta$ is the vertical mass flux normalized by the flux at the cloud base, and $E$ and $D$ represent entrainment and detrainment of mass from the cloud with the ambient atmosphere at each $z$. In RAS, the vertical continuity equation is simplified by assuming a quadratic profile \citep{Moorthi1999} and each cloud ``type" in the spectrum is characterized by the entrainment parameter $\lambda$ which specifies the mass entrainment per layer. Specifically, a cloud \texttt{i} with an entrainment rate $\lambda_i$ will entrain dry air throughout its ascent and become neutrally buoyant at layer \texttt{i} above the cloud base. The RAS module therefore solves for the total contribution for all such possible convective clouds within the column whose net convective energy matches the change in CAPE within that column from grid-scale dynamics. Finally, the grid-scale variables (moisture content, temperature, etc.) are updated based on the total contributions of each cloud type with a relaxation parameter $\tau_{relax}$, which is the characteristic timescale over which sub grid-scale effects contribute to grid-scale dynamics. This is not well known on Jupiter or on Earth \citep{Moorthi1992}, but studies of convective resolving simulations show that clouds relax over the timescale of a few hours \citep{Hueso2001}. Indeed \citet{Sankar2022} find that $\tau_{relax} \sim 1$ hour produces reasonably realistic convective signatures, which is what we use in our study.

In summary, for each latitude-longitude grid in EPIC, we calculate the total CAPE for each timestep, and determine the change in CAPE from the previous timestep, which we assume goes fully into convection. In theory, an increase in CAPE does not always lead to convective cloud formation and convection is instead triggered by an increase in CAPE over a critical value, which is a poorly known function of local weather conditions, and is empirically determined on Earth \citep{ArakawaSchubert1974}. On Jupiter, in the absence of such measurements, we assume that the critical value of CAPE is 0 (i.e., any small convection is sufficient to produce clouds). However, such low energy convective storms will have very negligible effects on the atmosphere and thus we expect that their effects will not play a large role in long-term dynamics investigated here. 

Note that unlike the diagnostic steps detailed in Section~\ref{sec:CAPE} above, in the RAS module, the change in CAPE is determined by the convolved effects of all the processes listed above. Specifically, the change in grid-scale moist static energy due to dynamics in each timestep happens simulatenously from all three processes and thus it is difficult to disentange the specific contribution from each process plays. As such, the RAS solves for the convective flux from these processes prognostically, without making the simplifying assumptions that we have made. We then use this diagnostic method to interpret how convection transpires in the model atmosphere by breaking down the role of each process.

\subsection{Equalizing the initial state}
Since the initial state is not balanced to floating-point precision and generally accrues integration errors, we initially run the model in a `2-D mode' (i.e., latitude vs. pressure) without cloud formation or moist convection for 200 Earth days. This is to ensure that the initial state stabilizes in a way that does not introduce any energy/vorticity/spurious cloud formation into the model atmosphere. During this phase, we found that the initial unstable state quickly advects vapor to unphysical locations due to cloud microphysics being disabled, and thus we add a relaxation term which adjusts the vapor profile for the condensing species to the initial state over a timescale of $5$ days. We tracked the amount of lost mass as a result of the relaxation and found that it is consistently under 1\% of the total initial vapor mass, but significantly improved the stability of the model when condensation was turned back on.  We chose 200 days, since most of zonal and meridional winds stop fluctuating after about 100-150 days, and we allow an additional 50 days to ensure better convergence. After this stabilization phase, we copy the 2D grid along the zonal direction, as the input for our 3D simulation with cloud physics.

\subsection{Perturbing the atmosphere}
At the end of the 2D phase, we are left with a perfect zonal symmetry, which is not representative of the real planet. To perturb the atmosphere, we introduce noise to our wind field through 100 small, randomly generated \paperedit{gaussian noise to the wind field}. The added wind speed is under 5 m/s to ensure that it does not significantly alter the vorticity and zonal wind profile of the planet, but is enough to produce a noticeable effect. These vortices are between 0.5 to 1$\degree$ across, 1 scale-height vertically and randomly positioned between 500~mb and 5~bar (i.e., the main weather layer). To model Jupiter's intrinsic luminosity, we also apply heating to the bottom atmosphere (below 20 bar) of $0.008$ W/kg and cool the top of the atmosphere (above 50 mb) at a rate of $0.01$ W/kg. This results in a net heat flux of about 7-10 W/m$^2$ across the planet, which is consistent with observations of Jupiter's internal luminosity \citep{Li2017}. We then run the model for 150 days with cloud microphysics enabled for water and ammonia, and with the RAS scheme enabled. 

\section{Results} \label{sec:results}
Out of the suite of $m$ values tested in the model we found that the $m=0.5$ is incredibly unstable and leads to severe issues with the model. Since this is consistent with the results of other studies which use this profile \citep[e.g.,][]{GarciaMelendo2005}, we are confident that this value of $m$ is likely inconsistent with the jovian atmosphere, and the true value of the wind shear slope is much lower. Therefore, for our remaining analysis, we showcase only the three values of $m=0$, $m=0.25$ and $m=0.4$.

\subsection{Cloud formation}\label{sec:results_cloudformation}
Firstly, we found that the introduction of the perturbations creates a chaotic environment for the first 20 days in the model which is unrepresentative of the long-term evolution of the atmosphere. Therefore, we focus primarily on the latter part of the simulations (day 50 - 150) in order to draw conclusions. 

\begin{figure*}
    \centering
    \includegraphics[width=\textwidth]{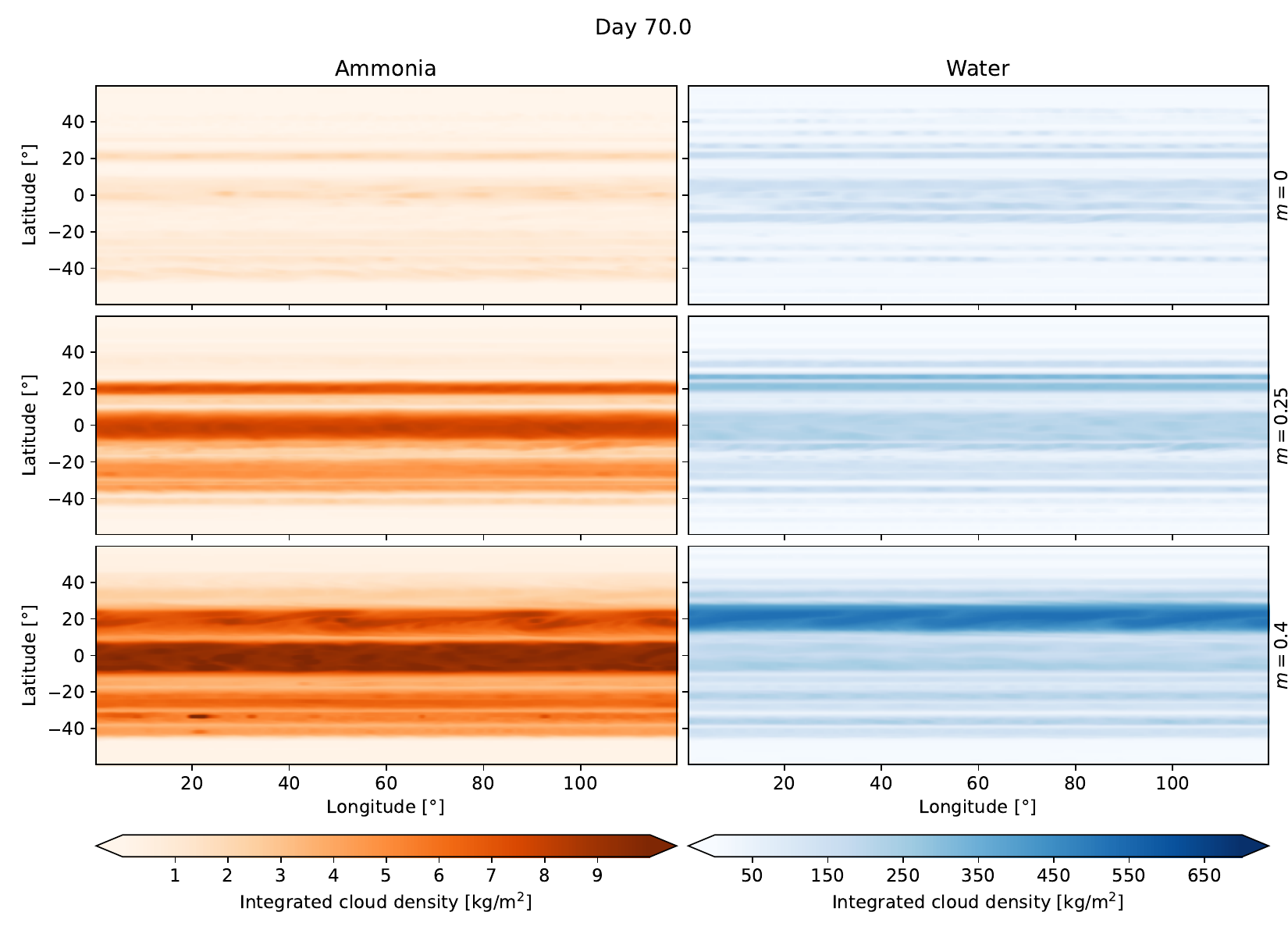}
    \caption{The vertically integrated ammonia (left) and water (right) clouds in the model 70 days into the simulation after adding the perturbations. Note the increase in both the ammonia and water cloud density with the increase in wind shear.}
    \label{fig:clouds}
\end{figure*}

Secondly, we find that generally, the creation of planetary scale \paperedit{vortex wave trains} is more prominent with increased wind shear. Figure~\ref{fig:clouds} shows the cloud structure in the model for our different cases at day 70, for water and ammonia clouds. For both clouds, we vertically integrate the cloud density, showing essentially a top-down view of the cloud structure. We see that the zero-shear case shows almost no zonal asymmetry (or longitudinal structure) in the ammonia, and increasing the wind shear leads to more turbulent features, particularly in the lower latitudes. 
For the water clouds, the zero shear case shows increased water cloud features in the lower latitudes, and southern hemisphere for the higher shear cases. There is very little cloud in the low shear cases. 
For both cloud layers, we see that there is an increase in cloud activity near the equator, and at specific latitude bands, but interestingly, the signatures in the water and ammonia cloud layers differ. 

\begin{figure*}
    \centering
    \includegraphics[width=\textwidth]{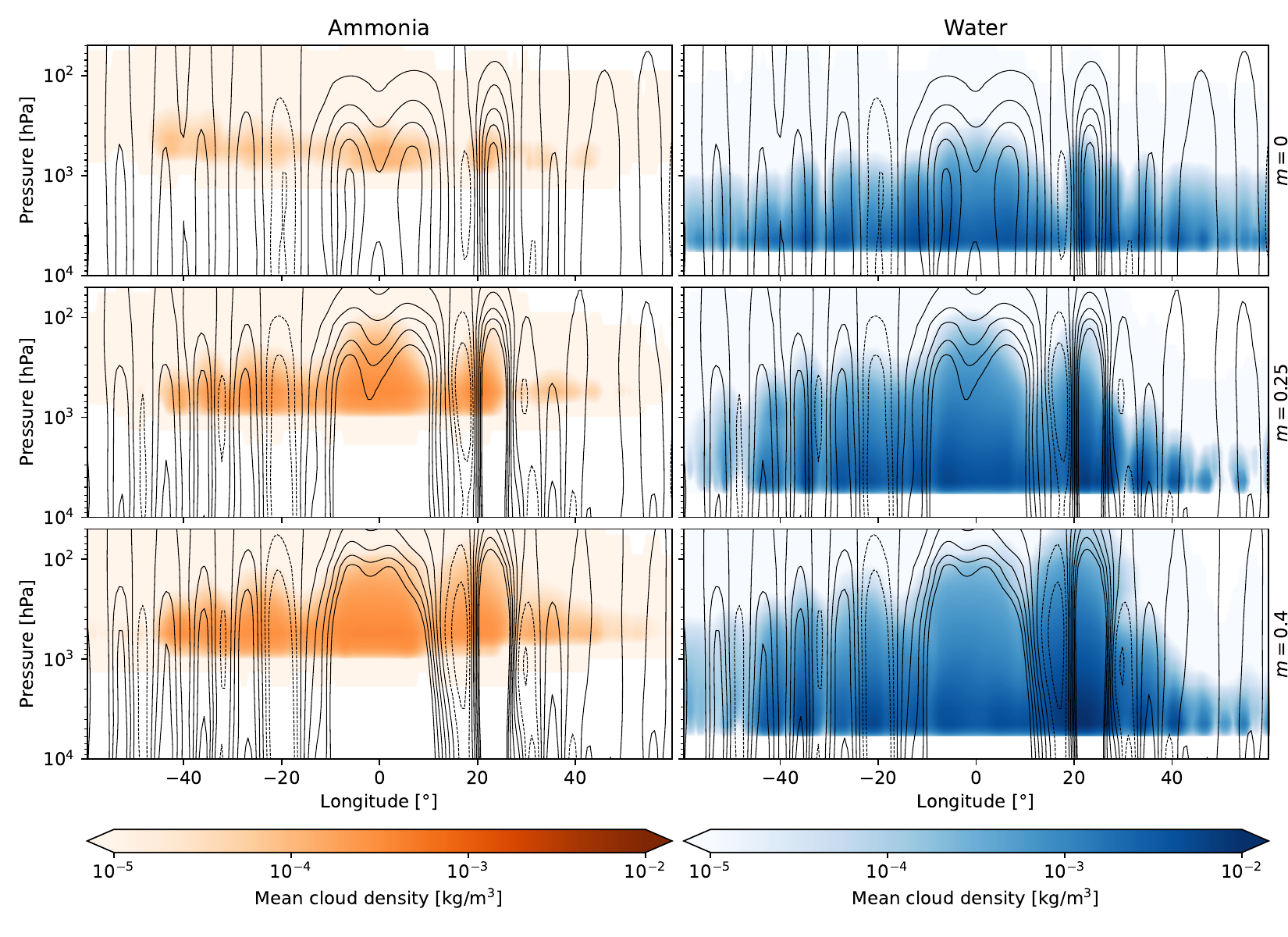}
    \caption{Zonally and temporally averaged ammonia (left) and water (right) cloud density for each of the cases studied. The zonal wind profile is shown with the black lines, with each contour separated by 25 m/s. We see that clouds are forming thicker higher in the atmosphere at specific locations due to convection carrying volatile rich air to the upper regions. These are prominently at high meridional shear or in the peaks of jet streams.}
    \label{fig:clouds_vertical}
\end{figure*}

Figure~\ref{fig:clouds_vertical} shows the zonally averaged cloud density at day \paperedit{70} for water and ammonia. There are clear morphological differences between the three simulations. Specifically, the higher shear cases show demonstrably more cloud formation near the equatorial region where convection drives water and ammonia to lower pressures. \paperedit{We found that cloud densities were two low compared with previous work, so we revised our density formulation to include both non-precipitating (i.e., cloud ice in EPIC) and precipitating particles (snow phase in EPIC). This more inclusive aerosol definition is more relevant to remote-sensing data which are sensitive to all aerosols at the moment of observation, and it produces a distribution that is more spatially homogeneous. They are now comparable to other studies of cloud formation, typically between $10^{-4}$ to $10^{-2}$ kg/m$^3$, which is an order of magnitude higher than previous studies of cloud formation \citep[e.g.,][]{Sugiyama2014}, but comparable to thick convective clouds on Earth \citep{Mansell2005} and cloud formation on Jupiter with strong updraft velocities \citep{Wong2015}.} Near the equator in our high-shear simulations, convection lifts volatiles up to just below the tropopause resulting in taller and thicker clouds at this location. This also occurs around $20\degree$ N, although storms here are not nearly as powerful. The equator and $20\degree$N are also where our methane-band scaling (Fig.~\ref{fig:clouds_vertical}) defines the highest cloud tops, providing a self-consistent justification for the scaling assumption.

\subsection{Moist convection intensity}
Due to the nature of the RAS scheme, it is difficult to get total convective \paperedit{energy and mass} fluxes since each convective trigger is variable with both the timestep and the parcel updraft profile. Therefore, we can instead diagnose convection by observing the effects of convection in the atmosphere. Specifically, we can track the amount of water vapor moved to the upper levels through convection, and plot the ratio of the vapor at these upper levels relative to a location near the cloud base (4 bars). In our following analysis we track the vapor at 600 mb and 1 bar, and then plot the ratio of these two levels with respect to the vapor at 4 bars.

\begin{figure}
    \centering
    \includegraphics[width=\columnwidth]{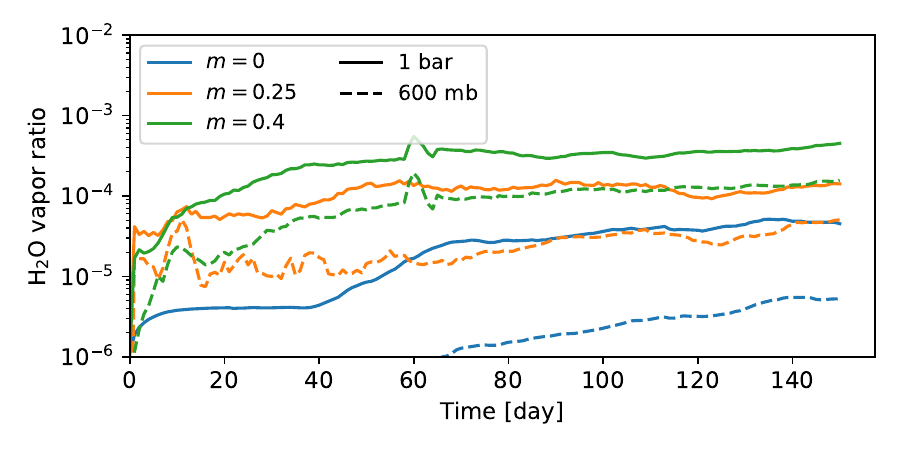}
    \caption{The ratio of water vapor mass mixing ratio between 1 bar and the cloud base at 4 bars (solid) and 600 mb and 4 bar (dashed), as a function of model time. The ratio is globally averaged for each timestep. We see that convection carries water to the upper altitudes, and is generally sustained throughout the duration of the model.}
    \label{fig:water_vapor_day}
\end{figure}

Figure~\ref{fig:water_vapor_day} shows the ratio of water vapor at 600 mb and 1 bar as a function of time. 
The response from simulations without wind shear is different from simulations with wind shear below the cloud tops: there is very little convection in the zero-shear case, while there is significant convection in both the $m=0.25$ and $m=0.4$ cases. 
Furthermore, we see that the convective flux tapers out after about 35-40 days (particularly for the deep convection that reaches 600 mb). Even though water vapor is sustained at 600mb through upwelling, the ratio stabilizes at a value of $\sim10^{-5}$ for both $m>0$ cases.
While moist convection efficiently transports vapor in the $m > 0$ cases, the $m = 0$ case has a much slower rate of vapor transport and does not reach an equilibrium state by the end of the 150-day simulation. The difference is primarily due to the limited vertical extent of moist convective activity in the $m = 0$ case, as reflected by the vertical distribution of water clouds in Fig.~\ref{fig:clouds_vertical}.

\begin{figure}
    \centering
    \includegraphics[width=\columnwidth]{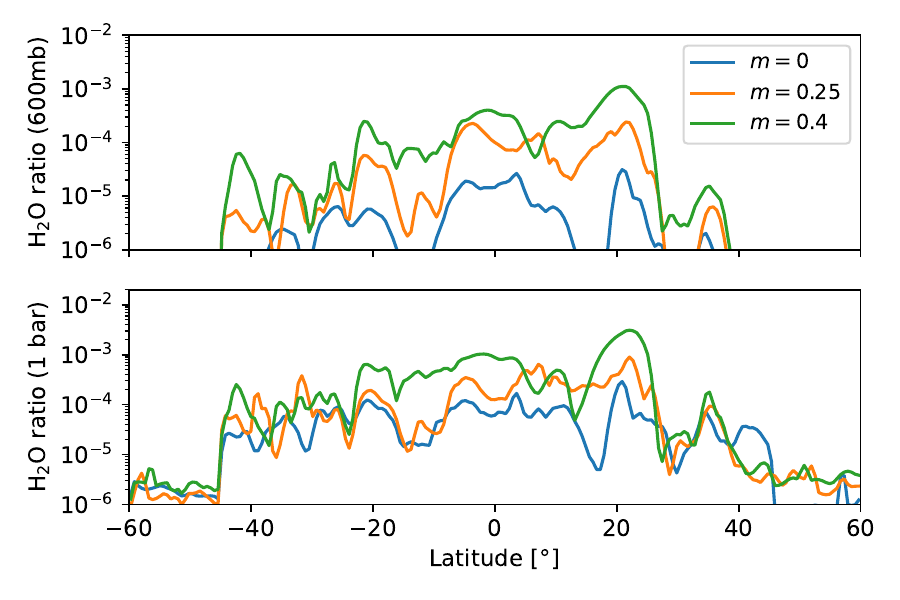}
    \caption{The ratio of water vapor mass mixing ratio between 600 mb and 4 bars (top) and 1 bar and 4 bar (bottom), as a function of latitude. Here, the ratios are zonally and temporally averaged for each simulation. We see that the ratio peaks at specific latitudes, which show dominant convective intensity. }
    \label{fig:water_vapor_ratio_lat}
\end{figure}

Figure~\ref{fig:water_vapor_ratio_lat} shows the water vapor mixing ratio at 600mb and 1 bar as a function of latitude. Here, the mixing ratio is temporally and zonally averaged showing, effectively, the convectively sustained water clouds in the upper atmosphere. There are a few observations to make. Firstly, as the wind shear increases, so too do the CAPE tendencies. Secondly, there is very little convection north of $24\degree$ in our model. Finally, we see that there are specific regions that show increased convective activity, and these coincide with the thicker clouds seen in Figure~\ref{fig:clouds_vertical}. 

To diagnose the source of the increase in CAPE, we calculate the values of the processes outlined in Section~\ref{sec:CAPE} above. To do this, we assume that the major change in each variable is from advection, and therefore, calculate the tendency purely using the material derivative, 
\begin{equation}
    \dfrac{\partial}{\partial t} = - u \dfrac{\partial}{\partial x} - v \dfrac{\partial}{\partial y},
\end{equation}
where $x$ and $y$ are the local length scales along the zonal and meridional axes, respectively, accounting for the oblateness of planet. \paperedit{The} vertical velocities \paperedit{in our model} are significantly smaller than the horizontal velocities and are therefore negligible for transport.

\begin{figure*}
    \centering
    \includegraphics[width=\textwidth]{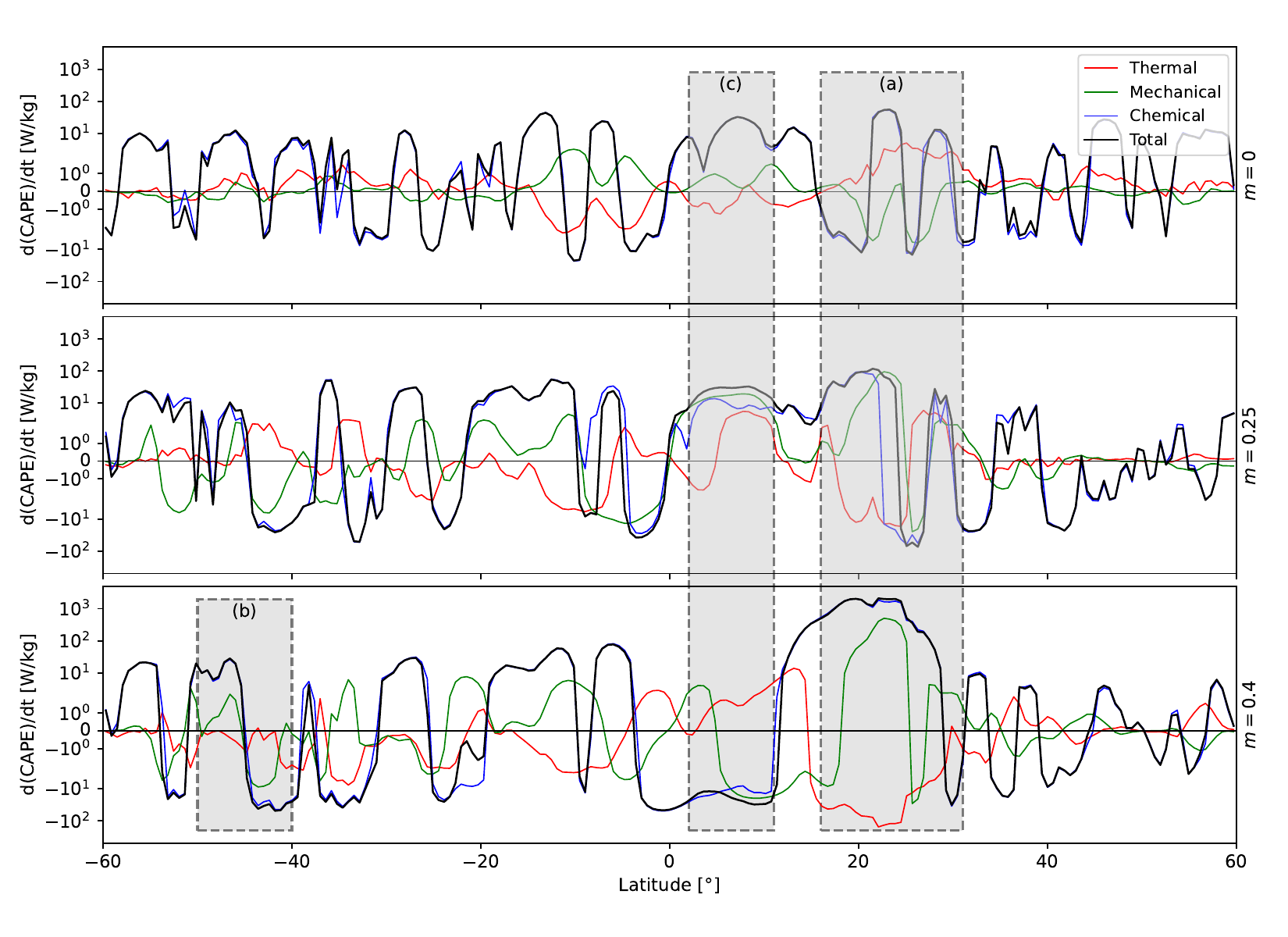}
    \caption{The contribution to the increase in CAPE from each of the processes described in Section~\ref{sec:CAPE}. The total value is shown in the black line. We can see that across all the cases, the ``chemical'' effect (i.e., advection of water vapor) is the most prominent driver of CAPE increase across all the processes. (a), (b) and (c) correspond to regions that show high convection across all cases, high convection in the high shear case and low convection, respectively.}
    \label{fig:convective_processes}
\end{figure*}

Figure~\ref{fig:convective_processes} shows the relative increase in CAPE from each of the three processes in Eqn.~\ref{eqn:decomposed}. Here, we see that the locations with peaks in the total CAPE tendency correspond well with the locations of convection in Figure~\ref{fig:water_vapor_ratio_lat}, as traced by the large increase in the vertical water vapor flux.

The scale of the water vapor term in the CAPE tendency also increases by an order of magnitude between the different case at some locations (see e.g., at $20\degree$ N and $45\degree$ S, given by boxes a and b respectively).
Here, the dominant process that drives CAPE is from increase in moisture at the cloud base, since the increase in moist static stability from an increase in vapor (blue line) is at least an order of magnitude larger than the other terms in the model.
In other locations, such as at around 5--7$\degree$N for all cases (box c), there is an increase in CAPE due to increase in moisture, but the other terms are sufficiently negative to counteract this increase. 

\subsection{Advection of water}
As detailed above, the convection in the atmosphere appears driven mainly by convergence of water vapor. However, the regions that demonstrate increased convection appear to have a spectrum of dynamics (e.g., some regions reside in strong meridional wind shear, while others are at the peaks of zonal jets). We investigate these regions in more detail below with relation to the local water vapor dynamics.

\begin{figure*}
    \centering
    \includegraphics[width=\textwidth]{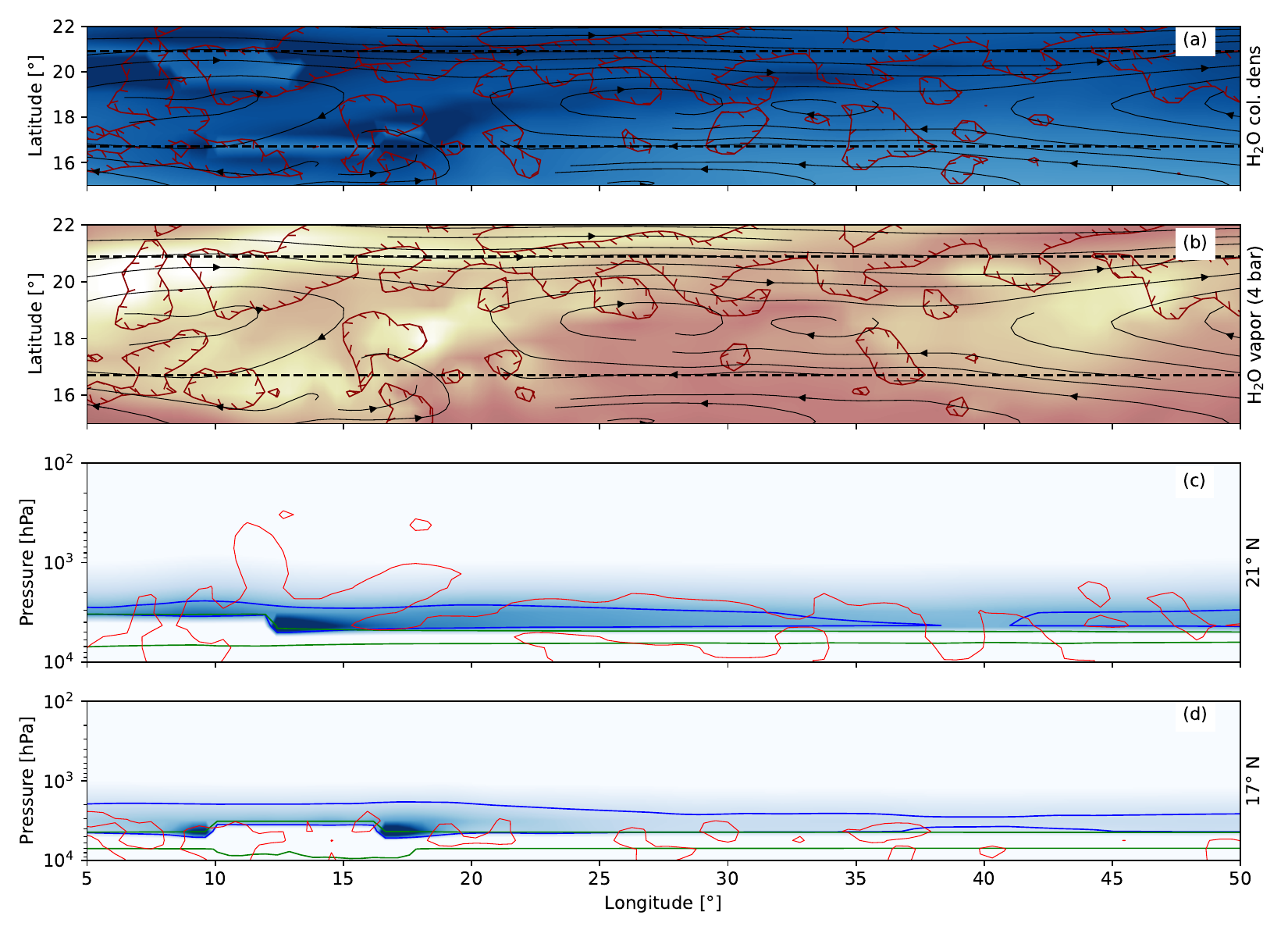}
    \caption{(a) Integrated water cloud density \paperedit{(with the same colorbar as in Figure~\ref{fig:clouds})}, with the local 4 bar wind field in black and the location of water vapor convergence (i.e., local increase in water vapor) in red. (b) 4 bar water vapor mixing ratio, on a linear scale ranging from $5\times10^{-4}$ (black) to $5\times10^{-3}$ kg/m$^3$ (white). (c,d) Cross-sectional view at 21$\degree$N and $16\degree$ N, respectively, showing the water cloud density as blue shaded regions \paperedit{(corresponding to water cloud density between $10^{-3}$ to $10^{-2}$ kg/m$^3$)}. This corresponds to the $m=0.25$ case at 110 days. The locations of the water vapor convergence is shown in red, and the local water snow and rain is shown in blue and green respectively. We see how the convective storms form upstream of the water vapor convergence, while precipitation happening behind the top of the tower (the bulge in the snow contours are to the left of the convergence at $21\degree$N and to the right at $17\degree$N). The eddy transport of water vapor is responsible for initiating and maintaining convection. The downdrafts from precipitation refuel the next eddy wave packet.}
    \label{fig:clouds_top_slice}
\end{figure*}

Figure~\ref{fig:clouds_top_slice} shows the top down view over the North Equatorial Belt with two zonal transects at $21\degree$ N and $16\degree$ N. The first two panels show the vertically integrated water cloud density and the 4 bar water vapor mass mixing ratio respectively, with the corresponding wind field in black, and areas with positive water vapor tendency in red, both at 4 bars. The last two show the zonal slice at $\sim 21\degree$ N and $\sim 16\degree$ N, respectively, as shown by the horizontal dashed lines in the first two panels. The water cloud density is shown as blue shading, snow is shown with the blue contours, rain with green and the water vapor tendency in red. 
At both these latitudes, we see that there are towering convective \paperedit{structures} which reach the upper troposphere \paperedit{(i.e., resulting in cloud formation at $\sim600$ mb)}.
Note how the increase in vapor happens at the leading edge of each convective plume (i.e., ahead of the convective tower with respect to the wind motion), showcasing how the advection of vapor leads to an increase in CAPE, which triggers convection and the maintenance of the convective tower.
In the trailing side of the tower, we see precipitation with the snow, which drops and becomes rain below the cloud base (where it is warm enough to melt). This recycles the vapor below the cloud base and triggers again through the convergence of water vapor. 

Specifically, we see from the cloud base water vapor mixing ratio that the convergence of water vapor is driven by the presence of a local horizontal circulation, which brings in vapor rich moisture to a relatively dry region (i.e., note how the red regions highlighting increasing vapor mixing ratio are specifically at boundaries between vapor rich and vapor poor areas). This influx of vapor through the local turbulence enriches the air, and leads to an increase in CAPE, which drives strong convection. 

\begin{figure*}
    \centering
    \includegraphics[width=\textwidth]{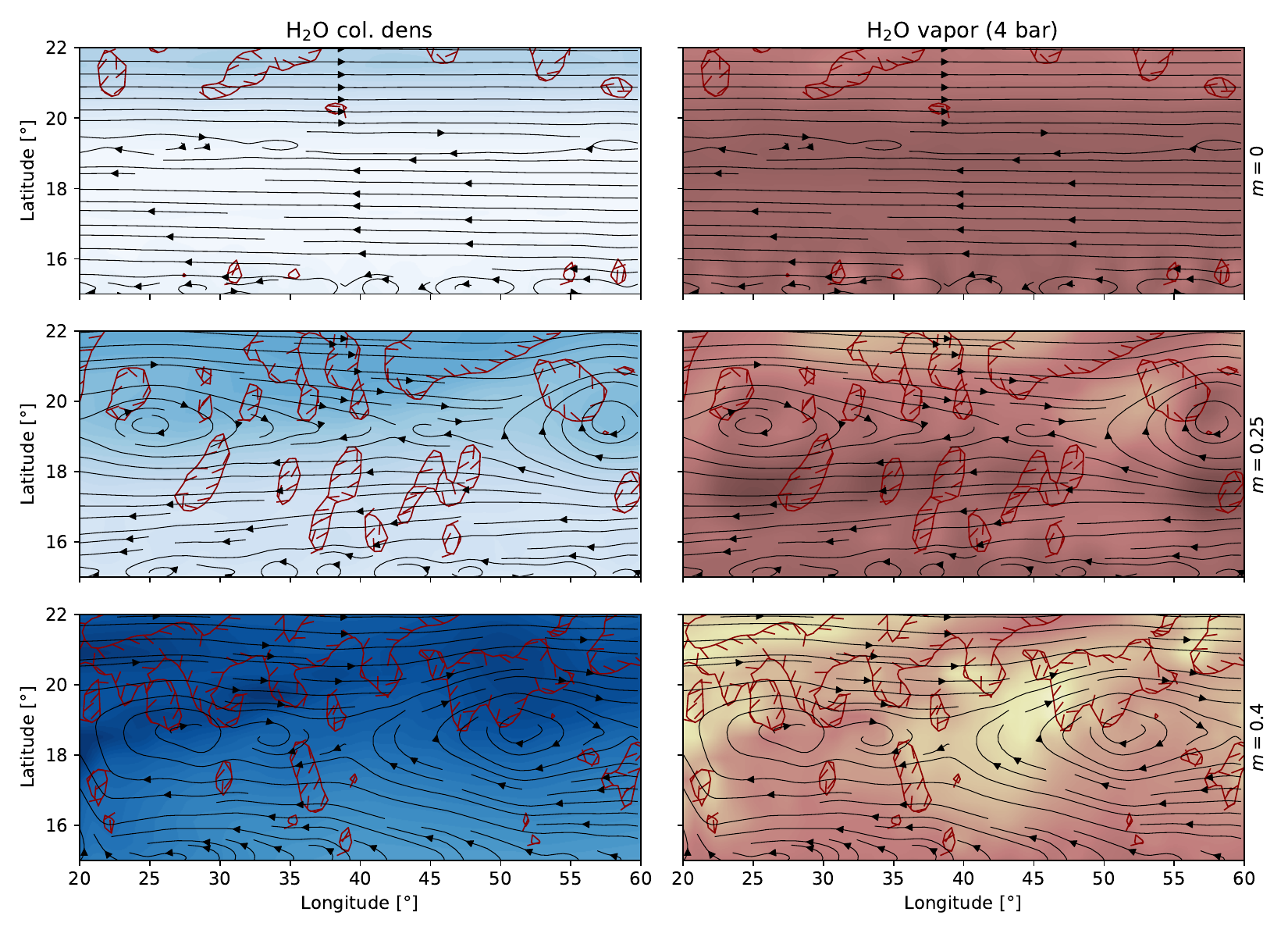}
    \caption{Integrated water cloud density (left, \paperedit{same colorbar as in Figure~\ref{fig:clouds}}) and water vapor mixing ratio at 4 bars (right), with the same colorscale as in Figure~\ref{fig:clouds_top_slice}, along with the associated wind field at 4 bar and the water vapor convergence in red, at day 110 into the model. Note how both the turbulence and the meridional volatile gradient increases with $m$, increasing the strength of convection in the atmosphere. }
    \label{fig:advection_cases}
\end{figure*}

Figure~\ref{fig:advection_cases} shows the change in the water vapor advection in this region with the different cases, and the difference in the CAPE is immediately evident. Specifically, the higher shear cases show more turbulent eddies at the cloud base, which mixes more water vapor meridionally. As such, locations where deep rooted eddies are more prominent drive convection in our model, and thus it is important to look into dynamical stability of the deep atmosphere. 

\subsection{Relation to the zonal wind profile}

\begin{figure*}
    \centering
    \includegraphics[width=\textwidth]{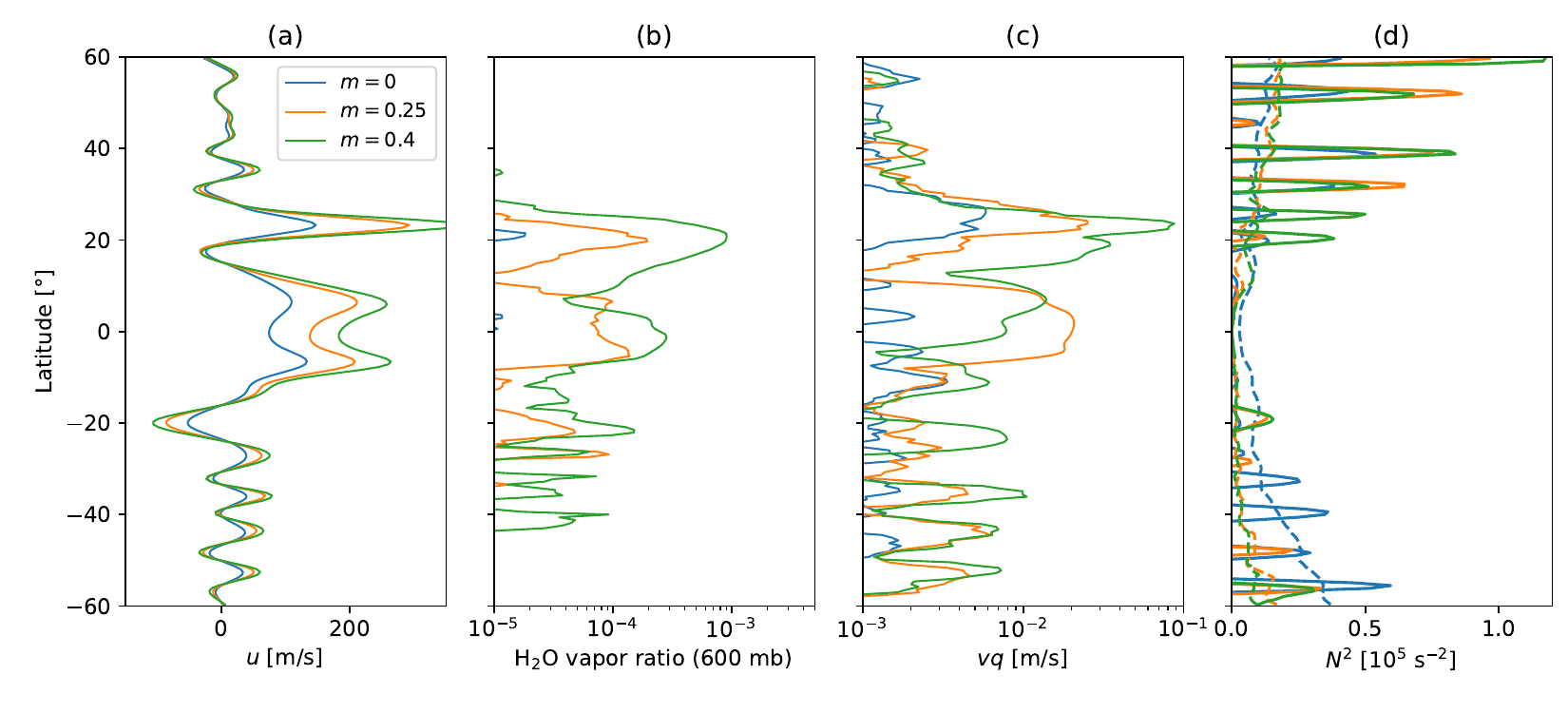}
    \caption{(a) The average zonal wind profile at 4 bars. (b) The temporally and zonally averaged ratio of water vapor mass mixing ratio between 600 mb and 4 bar, which denotes the convection due to water (c) the eddy meridional water vapor mixing at 4 bars. (d) the critical $N^2$ for instability formation based on the Eady model (solid) and the atmospheric $N^2$ at 4 bars (dashed). The locations of strong convection require both a strong meridional gradient in water vapor as well as strong eddy formation, resulting in water vapor mixing at the cloud base, highlighting the idea that the vapor transport is a primary driver for convection. \paperedit{The locations of these eddies correspond to locations where the atmospheric $N^2$ is smaller than the critical value required for instability formation in the Eady model. Note that the north/south boundaries of the model do not show strong convection even though the atmospheric $N^2$ is smaller than $N^2_{\rm crit}$ since the meridional vapor gradient is negligible, and thus even though there is eddy mixing, it does not lead to strong buildup of CAPE.}}
    \label{fig:convection_dynamics}
\end{figure*}

Figure~\ref{fig:convection_dynamics} compares latitudinal profiles of the deep zonal wind in the first panel, the vertical water vapor ratio (600 mb vs. 4 bars) in the second panel, \paperedit{the eddy vapor mixing ($vq$) in the third panel. In the fourth panel, we derive the critical Brunt-V\"ais\"al\"a frequency ($N^2$) for Eady instability \citep{Vallis2017} in solid line for each model and compare it with the model $N^2$ at 4 bars (dashed line) as a function of latitude. A model $N^2$ smaller than the critical $N^2$ is a necessary condition for instability formation and thus corresponds to locations which are likely sources of baroclinic instabilities}. Note how the both the meridional gradient of water vapor and $vq$ scale with wind shear, as the higher wind shear is increasingly unstable. An exception is the region north of $\sim24\degree$N, where $\sigma({\rm PV})$ is high for the low shear case, but zonal variability is suppressed for the $m$ = 0.25 and 0.4 cases. Generally, regions with high $vq$ (panel b) \paperedit{show more convection}. The $m=0$ case has minimal convection at most latitudes \paperedit{and a correspondingly minimum $vq$, except at around $20\degree$ N}. This is due to the fact that there is very little meridional gradient in the water vapor, and thus while eddies tend to form in the same location ($N^2 < N^2_{\rm crit}$), there is no discernable gain in CAPE since there is insufficient increase in water vapor mass from turbulent transport. For the other two cases, we see that CAPE is generated specifically through turbulent mixing of water vapor meridionally. 

\section{Discussions and Conclusions} \label{sec:discussion}
In this study, we have used the EPIC atmospheric model to study the global dynamics of deep water-based convection. We tested different vertical zonal wind shear profiles and found that there is a noticeable difference in the strength and location of convection in our model atmosphere. We attribute this convection to be due to the turbulent transport of water vapor near the cloud which results in an increase in CAPE due the latent heat effects and sparks large convective towers. This mechanism dominates all other mechanisms of CAPE generation in model (e.g., both thermal and mechanical effects), and the strength of the eddy transport increases with zonal wind shear. While this successfully explains the generation of convection in our model, the latitude distribution is noticeably different from the observed convection on Jupiter. We attribute this to several differences between our models and the real planet, as detailed below. 

\subsection{Shortcomings of the CAPE diagnosis}
We broke down the dynamical processes that affect the increase in CAPE to three terms: one which is purely thermal (i.e., advection of temperature), one which is `mechanical' (i.e., advection of thickness or vorticity) and finally the `chemical' term, which is the advection of water vapor to either increase its concentration (and CAPE) or decrease it and erode CAPE. While the RAS moist convective scheme is much more sophisticated in its treatment of buoyancy, we use several simplified assumptions in this diagnosis, which in some cases might over- or underestimate the change in buoyancy.

Particularly, we assume that there is little entrainment during the parcel ascent \paperedit{in our diagnostic breakdown of $d(CAPE)/dt$}. This allows us to simplify the vertical profile of the moist static energy by assigning it to the cloud base value. Ideally, accounting for entrainment should cause the moist static energy to decrease over the vertical extent of the profile ascent, and match the background saturated value at the cloud top \citep{ArakawaSchubert1974,Sankar2022}. While the entrainment profile is easy to calculate, it is much more difficult to diagnose the rate of change in the profile. However, for deep convection (i.e., with the cloud tops above $2$ bars), the rate of entrainment is fairly low \citep{Sankar2022}, due to the moist static energy in the upper regions being similar to the cloud base value. However, the convection in the RAS scheme calculates an explicit updraft profile \citep{Moorthi1999}, and therefore, the assumptions made here are valid for the conclusions drawn.

\paperedit{Moreover, while CAPE (and its variants that are used in Earth meteorology such as Most Unstable: MUCAPE or Surface-Based: SBCAPE, etc.) are useful quantities for diagnosing the potential for convective instability in the atmosphere \citep{Rochette1999}, they are not prognostic quantities in the convective energy budget of the atmosphere. Indeed, several Earth-based atmospheric models prefer to use the concept of the Cloud Work Function \citep[CWF][]{ArakawaSchubert1974, Arakawa2011}, which is the integral of the buoyancy weighted by the vertical mass flux. Here, the rate of change of the CWF provides a direct energy exchange between convection and horizontal kinetic energy in the atmosphere. This is much more difficult to do with the CAPE since the calculation of CAPE does not include the interaction between the convecting parcel and surrounding atmosphere. In Earth literature, the baseline CWF to trigger convection is an empirically determined quantity through in-situ measurements, which has no direct counterpart when applied to other planets. Some studies have attempted to derive vertical convective profiles from observations \citep[e.g.,][]{Hueso2022}, it is inherently difficult to derive 3D quantities from 2D observations. Therefore, while $d(CAPE)/dt$ (to first-order) describes the transfer of energy from the deep energy to the upper atmosphere through convection, the vertical distribution of energy through convection requires careful treatment of the vertical mass flux and other dynamical quantities. Understanding these relationships require more sophisticated models such as with those that resolve clouds \citep[e.g.,][]{Hueso2001,Li2019}.
}


\subsection{Distribution of lightning}
The observed distribution of lightning \citep{Brown2018} shows convection increasing in the belts and towards the poles. Indeed, lightning data shows almost no activity near the equator, except for a sharp peak around $8\degree$ N, which is similar to the location of plume activity seen in previous spacecraft data \citep{Smith1979}. Our model shows a similar peak in convective activity at $\sim 10\degree$ N (see Figure~\ref{fig:convection_dynamics}), but also significant, sustained convection near the equator. This is indeed a puzzling given both the ammonia distribution observed through radio observations show a peak near the equator \citep{Li2017,Moeckel2023}, as do other global simulations of convection on gas giants \citep{Young2019a,Young2019b}. However, \citeauthor{Young2019b} do note that the circulation cell that creates the increases the equatorial volatile distribution is likely not sustainable.

North of $24\degree$N, our high-shear cases show almost no convection, while night-side imaging from Galileo and Cassini show plentiful lightning in this region \citep{little++1999,Vasavada2005}, and Juno MWR and Waves instruments suggest that the northern high latitudes are the dominant source region for lightning signals \citep{kolmasova++2018,Brown2018}. Given the carrier of convection in our model, we interpret this to due to the lack of meridional variability in water vapor in the cloud base. However, note also that our model does not feature large persistent cyclonic vortices that are commonly seen on Jupiter at these latitudes (e.g., folded filamentary regions); FFRs are sites of frequent convective activity, as traced by their cloud structure, night-side lightning observations, clustered radio pulses seen by MWR, and even UV transient luminous events potentially analogous to terrestrial sprites \citep[see e.g.][]{Dyudina2004,Wong2020,imai++2020,giles++2020,wong++2023}. Our model does not contain persistent cyclonic vortices, which in observations can transition back and forth between convective and non-convective states \citep{Inurrigarro2020, Hueso2022}.

\subsection{`Moisture front' and eddy transport of water vapor}
\begin{figure}
    \centering
    \includegraphics[width=\columnwidth]{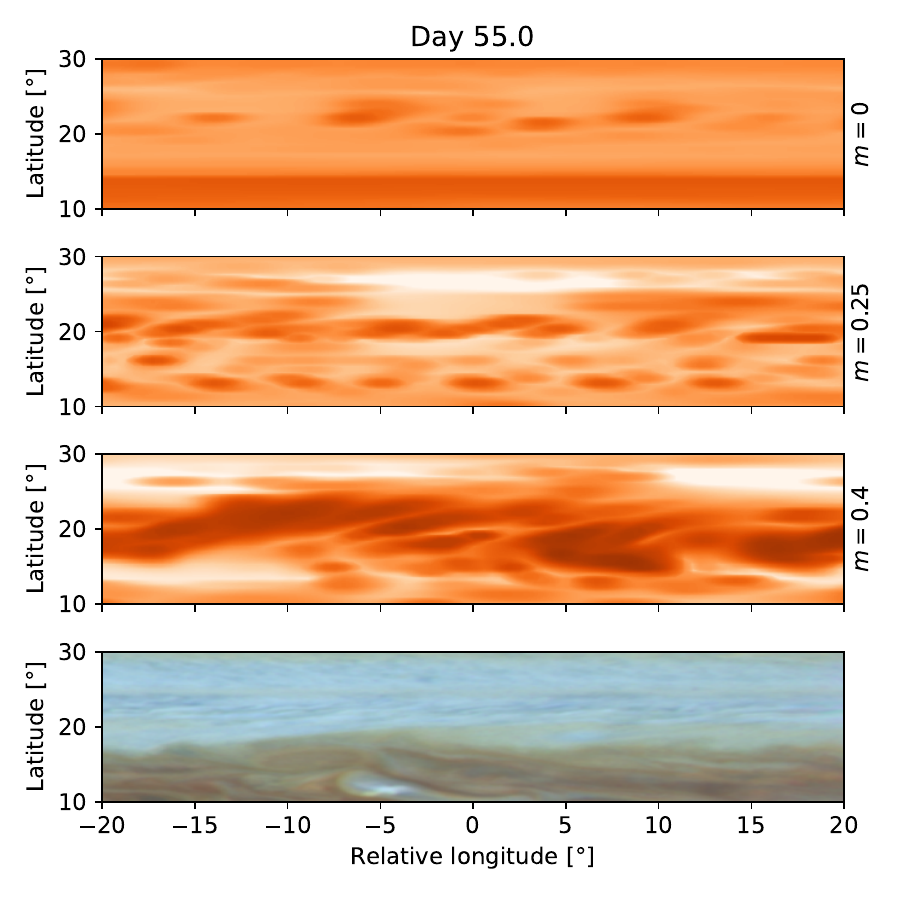}
    \caption{Integrated ammonia cloud column density for the different cases studied in the model, along with an Hubble OPAL \citep{simon++2015} image of the same region. Note the arc shapes in the ammonia clouds increase with wind shear, which matches the bright cloud patterns in OPAL observation. These clouds are set up by convective water clouds in our model, and the undulations are characterized by the deep eddy circulation. }
    \label{fig:ammonia_chevron}
\end{figure}

The primary driver for convection in our model is from the eddy transport of water. One primary example is the zone between $17\degree$N and $21\degree$N planetographic latitude (Fig.~\ref{fig:convection_dynamics}), where eddies mix volatile rich air from the north and south of the zone into the drier air, resulting in an increase in CAPE and convection. Lightning pulses detected by Juno MWR \citep{Brown2018} have local maxima primarily in belts with cyclonic wind shear, but this 
17--$21\degree$N zone
is a rare region with a local maximum in lightning frequency despite having anticyclonic shear typical of Jovian zones. Prior spacecraft observations did not detect night-side lightning flashes in this convective zone \citep{little++1999,Vasavada2005}, but this could be due to screening by the generally higher cloud opacity present in zones as compared to belts. Visible/IR cloud opacity does not attenuate the microwave pulses detected by Juno MWR. Our simulations in this region also reproduce numerous arc and chevron patterns that are seen in observations (see Figure~\ref{fig:ammonia_chevron}), which are also consistent with other models \citep{Young2019b}. 
Infrared observations show that this region does exhibit strong meridional gradients in volatiles \citep{Fletcher2020JGR}.

Juno observed an intense convective storm in the North Equatorial Belt, with Juno's MWR instrument revealing a strong meridional gradient in volatile concentration near the storm \citep{Brueshaber2023}.
These observations highlight why this region be conducive of the observed convective outbreaks, and further compound the idea of moisture advection fueling convecting events.

\begin{figure}
    \centering
    \includegraphics[width=\columnwidth]{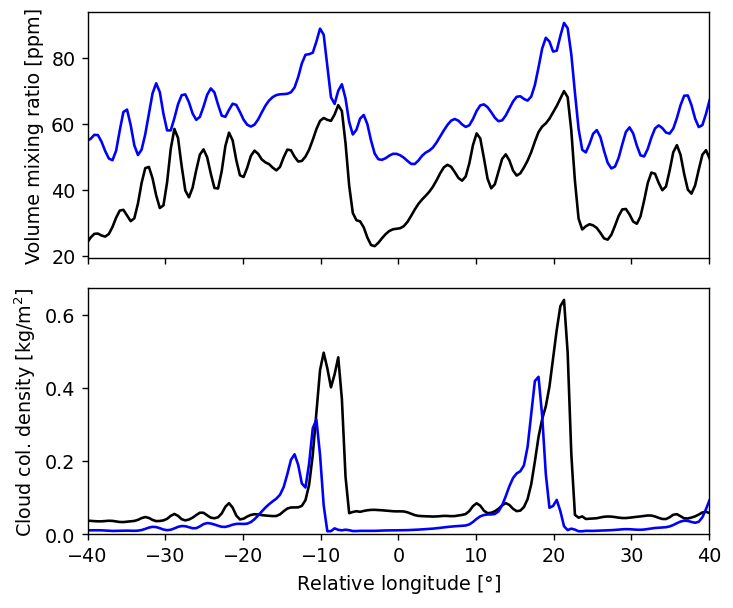}
    \caption{Top: Volume mixing ratio of water (black) and ammonia (blue) vapor near the 4 bar and 700 mb pressure level, respectively. Bottom: The integrated cloud density for water (black) and ammonia (blue). Both profiles are at $18\degree$ N latitude in our model for the $m=0.4$ case. Note how the thicker clouds for both species are coupled, due to being induced by convection by the water, as observed by \citet{bjoraker++2022}. However, both the ammonia clouds and peak in the ammonia vapor distribution are slightly downstream of the main convective tower for water due to the vertical wind shear.}
    \label{fig:volatile_ratio_slice}
\end{figure}

The vapor advection is an interesting process that is arguably difficult to observe. Specifically, while IR observations are able to retrieve ammonia abundances, these are generally degenerate with the atmospheric temperature and the pressure level sensed by the observations. 
\citet{bjoraker++2022} used high-resolution 5-$\mu$m spectroscopy to retrieve water and ammonia vapor mixing ratios in the 2--7 bar pressure range, measuring horizontal variation in these volatiles near hot spots and plumes in the equatorial region. While our model is unable to exactly recreate the periodicity and locations of these convective plumes, there is remarkable similarity in the zonal distribution of volatiles. \paperedit{The turbulent energy spectra of instabilities forming in shear-driven flows depends strongly on the magnitude of the shear \citep{Vallis2017} and thus it is possible that determining the periodicity of these features in our model and the real atmosphere differ due to the difference between the modelled and `real' shear, among others. Our expectation is that observations of this zonal periodicity can help constrain the vertical shear at these latitude bands.} Figure~\ref{fig:volatile_ratio_slice} shows our column-average ammonia and water mixing ratios, and corresponding vertically integrated cloud densities for the $m=0.4$ case. The wind blows eastward (to the right) in these zonal profiles at $18\degree$N. We see that the ammonia vapor and cloud lag behind the convective water plume (peaks in the water vapor and cloud distributions), bearing some similarity to the offset ammonia and water distributions in \citet{bjoraker++2022}. 

\begin{figure*}
    \centering
    \includegraphics[width=\textwidth]{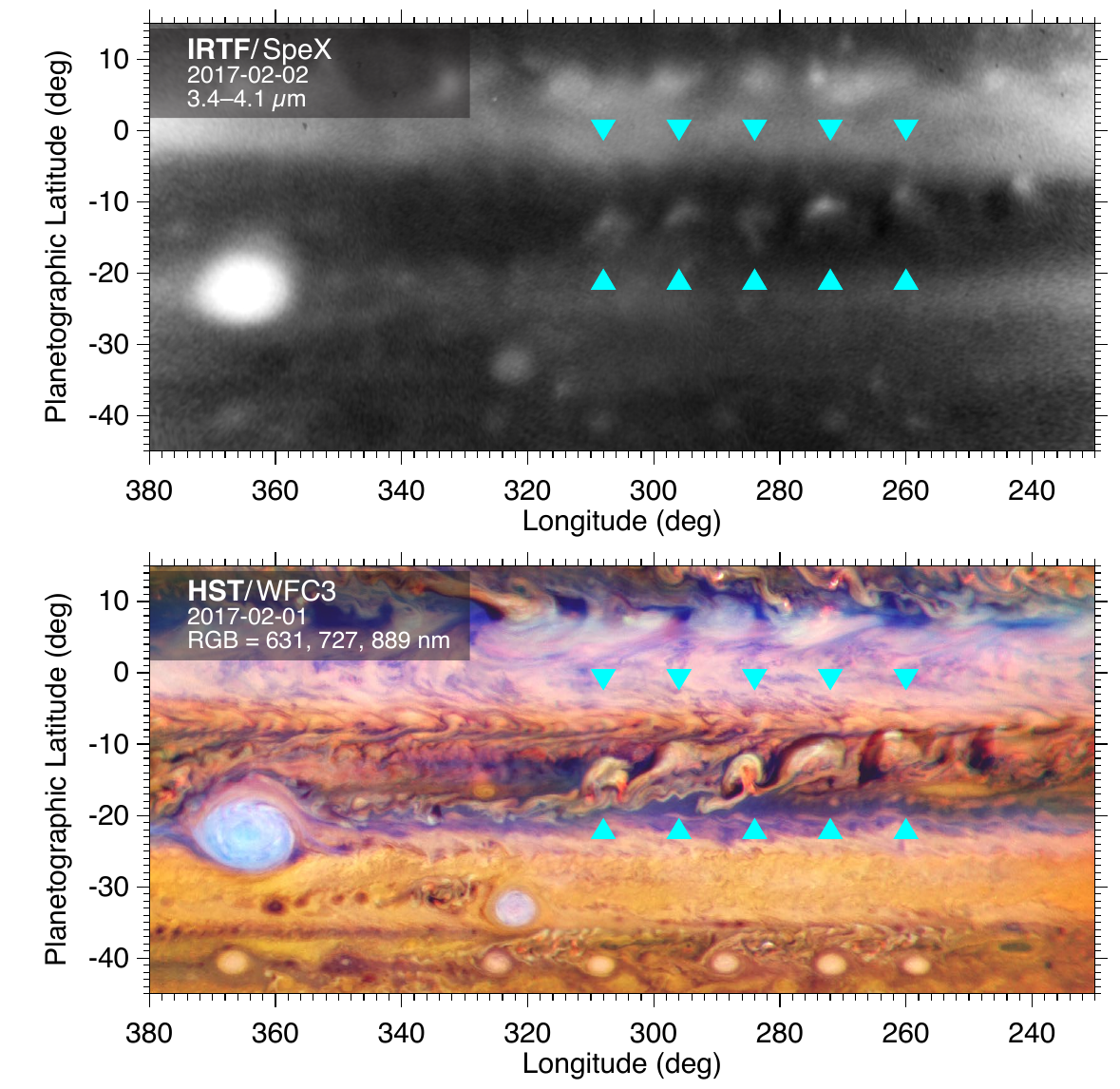}
    \caption{A periodic spacing of convective plumes was visible in the South Equatorial Belt in 2017, most clearly seen in the infrared at 4 $\mu$m (upper panel) where reflectivity is greatest from cloud/haze features present at high altitude or containing large particle (or both). The lower panel is a methane-band composite, where deep clouds appear red and high, clearings appear dark, and thick clouds appear white. The combination of these three elements is a signature of active moist convection \citep{imai++2020,wong++2023}. Blue arrows are evenly spaced 8$\degree$ longitude apart. IRTF data are described in \citet{orton++2017}, and HST data are described in \citet{Wong2020} and available from \citet{wong2017}.}
    \label{fig:stormtrain_pj4}
\end{figure*}

\subsection{Zonal shear profile and baroclinic instability}
Our model shows a significant effect on convection from the zonal wind shear. In some cases, we see that this effect is from the increase in the strength of turbulent eddy formation from increasing the zonal wind shear (see Figure~\ref{fig:convection_dynamics} at around $20\degree$N and $20\degree$S). This can be understood in terms of baroclinic theory, where increasing the wind shear creates a steeper meridional temperature gradient from the thermal wind equation. Increasing the wind shear makes the fluid increasingly unstable to small perturbations \citep{Holtonbook}. As such, in our models, we definitely find that increasing the shear amplifies the intensity and frequencies of eddy formation, particularly due to the combination of the initial perturbation and the heating applied. However, as noted above, we find that the generation of eddies does not directly translate to an increase in convection throughout the atmosphere, but instead amplifies convection at specific locations. This also tends to happen at locations of strong jet peaks, whereas the increase in eddy formation between the different deep wind shear profile is less profound in other locations (e.g., south of $30\degree$N), where the zonal wind speeds are much smaller.

\begin{figure}
    \centering
    \includegraphics[width=\columnwidth]{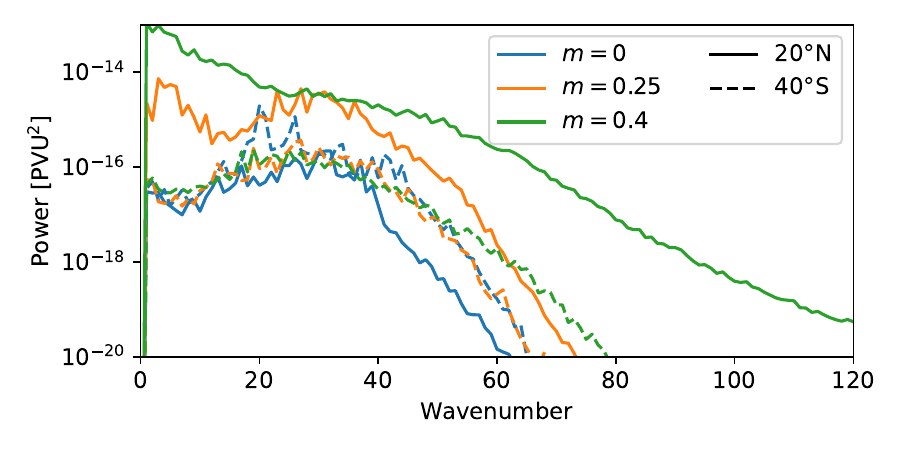}
    \caption{Spectrogram showing the distribution of wavenumbers for different shear cases, averaged over the last 100 days at a latitude of $20\degree$N and $40\degree$S. Note the difference in the spectral distribution at the two locations, and for $20\degree$N, how the higher shear cases show both an increased spectral power distribution across the wavenumber space and more eddies with higher wavenumbers (i.e., smaller longitudinal extent).}
    \label{fig:wavenumber}
\end{figure}

An interesting artifact of this process is that the turbulent eddies generate moving wave trains in our model, which generates periodic convection.  
\paperedit{These wave trains are a result of fluid dynamical instabilities which have strong correlation with the static stability of the atmosphere and the wind shear \citep{Vallis2017}}
\paperedit{In our model, } the amplitude and the wavenumber of the train is a function of the location on the planet and in some cases also a function of the shear, with larger shears producing more tightly packed wavetrains at some latitudes (i.e., higher wavenumbers, see Figure~\ref{fig:wavenumber}). \paperedit{This is expected since higher wind shear promotes growth rates of higher wavenumber waves, at least in a 2-layer geostrophic model \citep{Holtonbook}. While a full 3D treatment is complicated, we expect the same principles to apply.}
\paperedit{In our model,} the waves themselves exist deep below the water clouds driving deep horizontal circulation of volatiles, and the interactions between the crests of the waves which coincides with a meridional gradient in water vapor generates convection (e.g., an anti-cyclonic wave packet in zones leads to convection). 
These trains  were indeed thought to be a dominant source of convection in the equatorial region based on Voyager data \citep{Smith1979}. 
More recently, a periodic spacing  between individual storm plumes was observed in the South Equatorial Belt in 2017 by Juno, HST, Keck, and ALMA \citep{depater++2019,Wong2020,bjoraker++2022}. 
The individual plumes making up this superstorm eruption event were separated by about 8$\degree$ longitude (wavenumber $\sim$45), as revealed particularly clearly in infrared imaging near 4 $\mu$m (Fig.~\ref{fig:stormtrain_pj4})

In places where we observe strong meridional volatile gradients, but very weak convection, we should  expect very stable flows, which only weakly amplify baroclinic instabilities. Indeed, perhaps these locations, such as the NTrZ are likely to be significantly more barotropic with a very weak wind shear, given the lack of convection despite the strong meridional gradients in temperature \citep{Gierasch1986,Flasar2004,Sankar2021}. Locations with strong volatile gradients, such as the southern boundary of the NEB at $\sim 8\degree$N \citep{Li2017,Li2020,bjoraker++2022} 
or the southern boundary of the SEB at $20\degree$S, might be more susceptible to convection generated by baroclinic instability than the rest of the planet \citep{Brueshaber2023}. 

In the same vein, it is still unclear as to why our model shows very little convection in the northern mid- to high-latitudes, where Juno detects a greater frequency of radio pulses from lightning \citep{kolmasova++2018,Brown2018}. One possibility is that our initialization of the initial temperature field is strongly affected by the $24\degree$ N jet, through the integration of the thermal wind balance, and therefore results in cooler temperatures to the north of the jet. Combined with the cold-trap model to initialize the volatile distribution, it is possible that we over-estimate the wind shear of the jet, which overestimates the cold temperatures in these regions, and results in less volatile mass available for convection.

Indeed, we find that the 3D wind profile controls the nature of both the turbulent eddy formation as well as the meridional gradient in vapor, both of which are vital for the formation of convection in our model. Changing the 3D wind field (which in our model is done by changing the deep wind shear) affects the characteristics of turbulent eddy formation differently in different latitudinal bands. Specifically, note how the $m=0$ case shows increased turbulent eddy formation north of the $24\degree$N jet, while this is surprisingly lacking for the $m>0$ cases. Conversely, the models show increased turbulent eddies (with higher wavenumbers) south of this jet. 
The development of new observational constraints on Jupiter's deep wind field  (e.g., $m$ as a function of latitude), 
coupled with better constrains on either the temperature structure or the water abundance at depth, would steer future simulations toward more accurate models, and  improve our understanding of the nature of convection on Jupiter.

\subsection{Conclusions and Future work}
In summary, we studied convection in the jovian atmosphere using the EPIC GCM, and how the convective strength and intensity varies with deep wind shear. Our models show that a dominant carrier of convection is the advection of moisture-rich vapor near the water cloud base through eddy mixing from the formation of baroclinic instabilities, which are themselves increased in intensity with stronger vertical wind shear. However, we note that while the eddy mixing is stronger with increased wind shear, we also require a strong meridional gradient in water vapor to fuel convective storms. In our model, this gradient is an effect of the strong meridional temperature contrast across jet peaks, where the vertical wind shear results in a strong temperature gradient through the thermal wind balance. However, distributions of volatiles on Jupiter are generally much more inhomogeneous \citep{Li2017,Moeckel2023}, and a comprehensive study of the effect of the distribution of volatiles (particularly water) as it pertains to where and how convection form is a much more complex task, due to the lack of observations of the global distribution of water. This is out of the focus of our current study, but will investigate these effects in future simulations. \paperedit{Furthermore, we have used a constant wind shear with latitude, which affects the local cloud structure and eddy wavenumbers. Investigating the effect of varying the wind shear meridionally in an effort to constrain the wind shear profile with observed structures of eddies would be a valuable next step to pursue.}

\begin{acknowledgments}

The authors are grateful to G.S. Orton, for acquiring the data shown in Fig.~\ref{fig:stormtrain_pj4} under IRTF program 2017A034, and to the NASA/IPAC Infrared Science Archive (\url{https://irsa.ipac.caltech.edu}) for making the data available. HST data in Fig.~\ref{fig:stormtrain_pj4} were acquired under program GO-14661  and are available from the MAST Archive \citep{wong2017}. This work was supported by NASA Solar Systems Workings Grant \#80NSSC22K0804. Resources supporting this work were provided by the NASA High-End Computing (HEC) Program through the NASA Advanced Supercomputing (NAS) Division at Ames Research Center. We thank both anonymous reviewers for their thoughtful comments which have improved the clarity of this manuscript.

\end{acknowledgments}

%

\vspace{5mm}





\appendix

\section{Long term evolution and sensitivity to hyperparameters}
\begin{figure}[h]
    \centering
    \includegraphics[width=\linewidth]{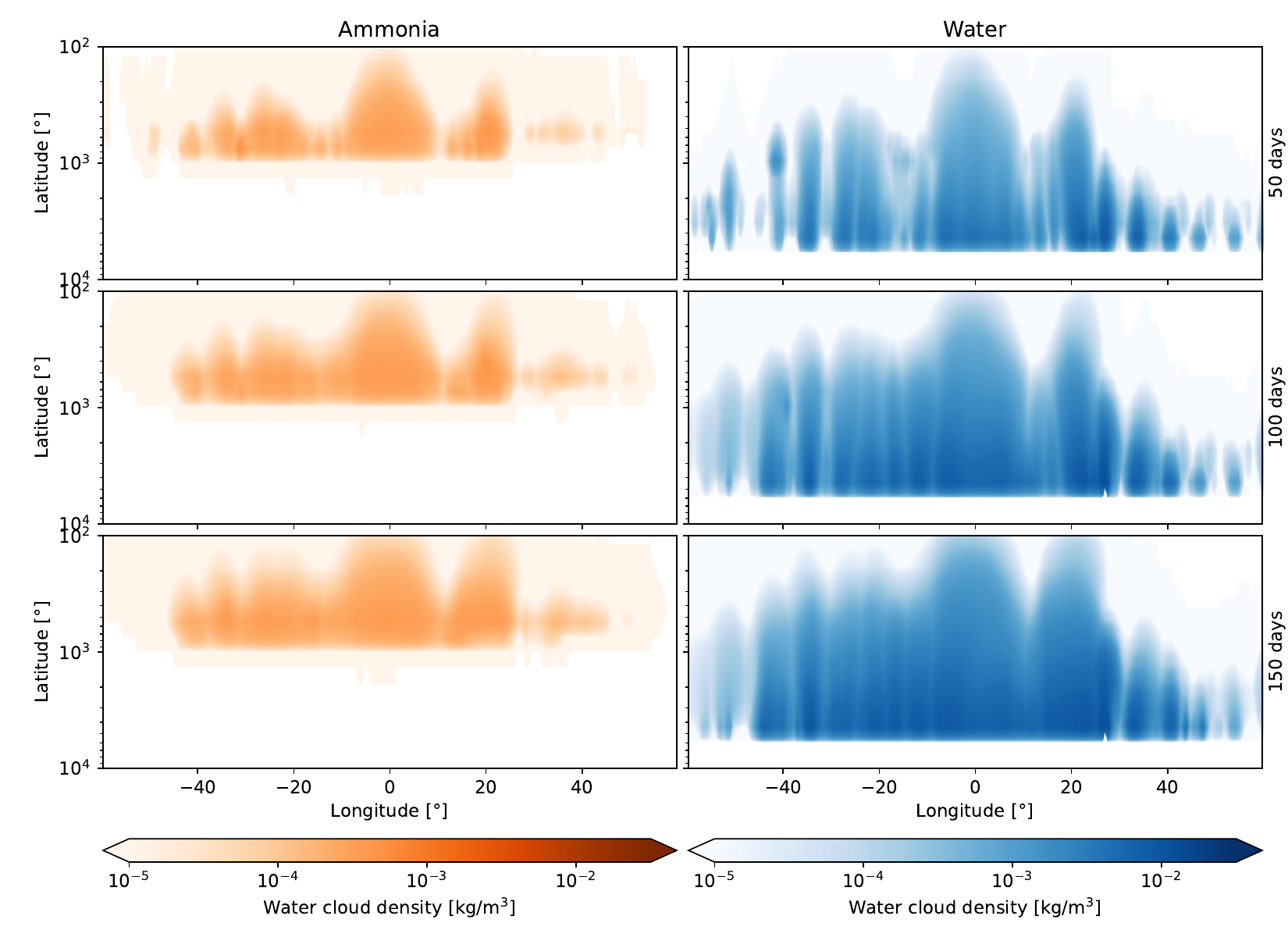}
    \caption{The zonal mean cloud densities at 3 different timesteps from our model atmosphere. There is little change in the cloud densities between the three timesteps showing that the model has reached a quasi-steady state by day 50, but cloud densities do still continue to evolve slowly during the latter part of the simulations.}
    \label{fig:cloud_evolution}
\end{figure}

To ensure our model reached a sufficient measure of steady state, we tested key atmospheric variables in our model atmosphere to ensure that they did not change significantly over the duration of the model. Figure~\ref{fig:cloud_evolution} shows the zonally averaged cloud structure at 3 different timesteps into the simulation for $m=0.25$.
We also extend the model to an additional 50 days and find that there is no appreciable change to the trends that we observe in the first 50 days.

\begin{figure}
    \centering
    \includegraphics[width=\linewidth]{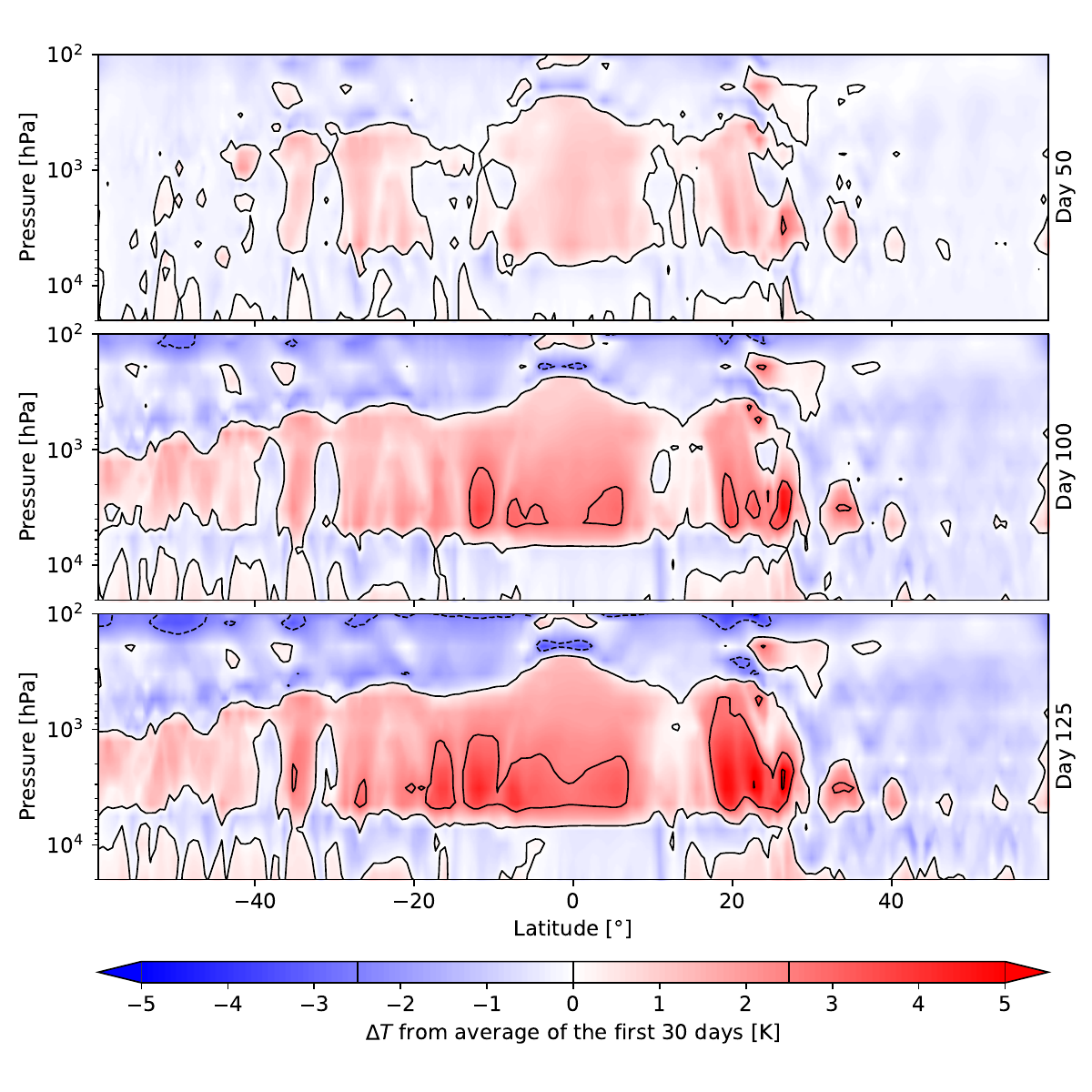}
    \caption{The difference between the temperature at days 50, 100 and 125 from the average temperature from the first 30 days of the simulation. The differences are zonally averaged. There is some evolution in the temperature field in the first 100 days, but the model achieves a quasi-steady state at this point after which there is a much smaller change in the temperature field.}
    \label{fig:temperature_evolution}
\end{figure}

Figure~\ref{fig:temperature_evolution} shows the difference of the zonally-averaged model temperature at day 50, 100 and 125 from the average of the first 30 days (i.e., the evolution of the temperature field from the initial temperature). The major temperature change is within the water and ammonia cloud level, and peaks at $\sim5$ K. The temperature and cloud fields do continue to evolve since we continue to apply a thermal forcing to the top and the bottom of the atmosphere.

We also vary specific model parameters which vary the stability of the model. We found that the model timestep produced the greatest change in the model stability. We tested $\Delta t=15, 30, 36$ and $60$ s and found that a maximum of $\Delta t \leq 30$ s provided the most optimal stability for all simulations. We therefore used $\Delta t = 30$ s for all our simulations.

\begin{figure}
    \centering
    \includegraphics[width=\linewidth]{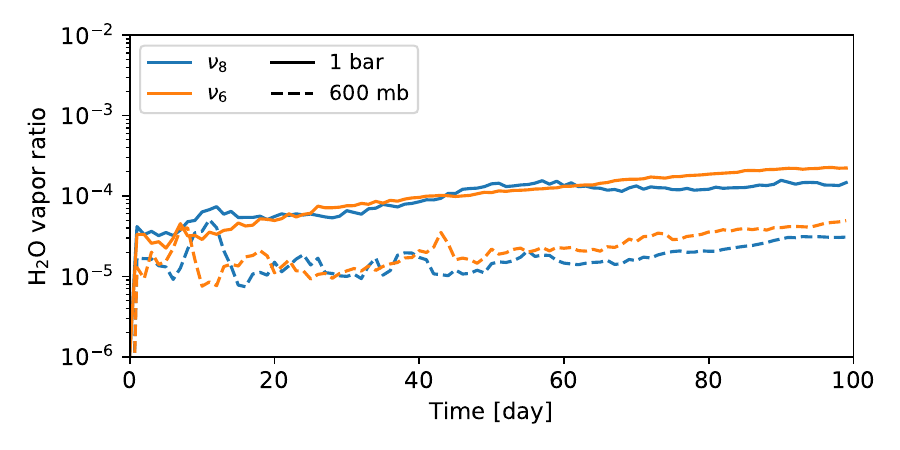}
    \caption{Effect of the model hyperviscosity on the convected water vapor. We find that the most prominent effect is in the first 10-15 days, whereas beyond this adjustment phase, there is negligible effect of the hyperviscosity on the model atmosphere.}
    \label{fig:hyperviscosity}
\end{figure}

Figure~\ref{fig:hyperviscosity} shows the effect of the model hyperviscosity on the resulting convective water vapor. We find that 
in the first 20 days there are fluctuations in the models with similar amplitude independent of the hyperviscosity, while the duration of the fluctuations is shorter for lower-order hyperviscosity. Beyond this 20-day adjustment stage, the model output is largely insensitive to the hyperviscosity order. In different wind shear environments, initial instability with higher-order hyperviscosity crashes the model, while lower-order hyperviscosity allows the model to recover from these instabilities, perhaps because the faster instability timescale limits feedback. 

\bibliography{ref}{}
\bibliographystyle{aasjournal}



\end{document}